\documentclass[12pt]{article}
\usepackage{fancyhdr}
\usepackage{authblk}
\usepackage{feynmf}
\usepackage{amsmath,amssymb}
\usepackage{graphicx}
\usepackage{amssymb}
\usepackage{url,color}
\usepackage{epstopdf}
\usepackage{caption}
\usepackage{subfigure}
\usepackage[colorlinks=true, citecolor=green, urlcolor=cyan]{hyperref} 
\def\stacksymbols #1#2#3#4{\def\theguybelow{#2}
    \def\vp{\lower#3pt}
    \def\sp{\baselineskip0pt\lineskip#4pt}
    \mathrel{\mathpalette\intermediary#1}}

\def\intermediary#1#2{\vp\vbox{\sp
     \everycr={}\tabskip0pt
     \halign{$\mathsurround0pt#1\hfil##\hfil$\crcr#2\crcr
              \theguybelow\crcr}}}

\def\beq{\begin{equation}}
\def\eeq#1{\label{#1}\end{equation}}
\def\eeqn{\end{equation}}
\newcommand{\gsim}{\hbox{ \raise3pt\hbox to 0pt{$>$}\raise-3pt\hbox{$\sim$} }}
\newcommand{\lsim}{\hbox{ \raise3pt\hbox to 0pt{$<$}\raise-3pt\hbox{$\sim$} }}

\newenvironment{Eqnarray}%
   {\arraycolsep 0.14em\begin{eqnarray}}{\end{eqnarray}}
\def\beqa{\begin{Eqnarray}}
\def\eeqa#1{\label{#1}\end{Eqnarray}}
\def\eeqan{\end{Eqnarray}}

\def\invfb{ \mbox{fb}^{-1} }

\def\roots{ \sqrt{s} }
\def\TeV{ \mbox{TeV}}
\def\GeV{ \mbox{GeV}}
\def\MeV{ \mbox{MeV}}

\def\be{\begin{equation}}  
\def\ee{\end{equation}}  
\def\bea{\begin{eqnarray}}  
\def\eea{\end{eqnarray}}

\providecommand{\ee}   {\rm{e^+e^-}}

\newcommand{\ttbar}{ t \bar t}

\newcommand{\eplus}{e^+}
\newcommand{\eminus}{e^-}
\newcommand{\epem}{\eplus\eminus}
\def\invfb{ \mbox{fb}^{-1} } 

\newcommand{\afb}{A_{FB}}
\newcommand{\afbt}{A^t_{FB}}
\newcommand{\thel}{\theta_{hel}}
\newcommand{\cthel}{\mathrm{cos} \theta_{hel}}
\newcommand{\qq}{q\bar{q}}
\newcommand{\Zzero}{Z^0}
\newcommand{\tpq}{t}
\newcommand{\Wboson}{W}
\newcommand{\bottom}{b}
\newcommand{\quark}{q}

\newcommand{\leftp}{\left(}
\newcommand{\rightp}{\right)}
\newcommand{\lhel}{\lambda_t}
\newcommand{\fonevI}{{\cal F}^{I}_{1V}}
\newcommand{\ftwovI}{{\cal F}^{I}_{2V}}
\newcommand{\foneaI}{{\cal F}^{I'}_{1A}}
\newcommand{\ftwoaI}{{\cal F}^{I}_{2A}}
\newcommand{\pem} { {\cal P} }
\newcommand{\pep} {{\cal P'} }
\let\bar=\overbar

\def\lsim{\mathrel{\raise.3ex\hbox{$<$\kern-.75em\lower1ex\hbox{$\sim$}}}}
\def\gsim{\mathrel{\raise.3ex\hbox{$>$\kern-.75em\lower1ex\hbox{$\sim$}}}}

\def\del{\partial}
\def\Dslash{\not{\hbox{\kern-4pt $D$}}}
\def\dslash{\not{\hbox{\kern-2pt $\del$}}}

\def\ee{e^+e^-}

\def\msb{{\bar{\scriptsize M \kern -1pt S}}}

\def\drb{{\bar{\scriptsize D \kern -1pt R}}}

\makeatletter
\def\section{\@startsection{section}{0}{\z@}{5.5ex plus .5ex minus
 1.5ex}{2.3ex plus .2ex}{\large\bf}}
\def\subsection{\@startsection{subsection}{1}{\z@}{3.5ex plus .5ex minus
 1.5ex}{1.3ex plus .2ex}{\normalsize\bf}}
\def\subsubsection{\@startsection{subsubsection}{2}{\z@}{-3.5ex plus
-1ex minus  -.2ex}{2.3ex plus .2ex}{\normalsize\sl}}

\renewcommand{\@makecaption}[2]{%
   \vskip 10pt
   \setbox\@tempboxa\hbox{\small #1: #2}
   \ifdim \wd\@tempboxa >\hsize     
       \small #1: #2\par          
     \else                        
       \hbox to\hsize{\hfil\box\@tempboxa\hfil}
   \fi}

 \def\citenum#1{{\def\@cite##1##2{##1}\cite{#1}}}
 
\newcount\@tempcntc
\def\@citex[#1]#2{\if@filesw\immediate\write\@auxout{\string\citation{#2}}\fi
  \@tempcnta\z@\@tempcntb\m@ne\def\@citea{}\@cite{\@for\@citeb:=#2\do
    {\@ifundefined
       {b@\@citeb}{\@citeo\@tempcntb\m@ne\@citea\def\@citea{,}{\bf ?}\@warning
       {Citation `\@citeb' on page \thepage \space undefined}}%
    {\setbox\z@\hbox{\global\@tempcntc0\csname b@\@citeb\endcsname\relax}%
     \ifnum\@tempcntc=\z@ \@citeo\@tempcntb\m@ne
       \@citea\def\@citea{,}\hbox{\csname b@\@citeb\endcsname}%
     \else
      \advance\@tempcntb\@ne
      \ifnum\@tempcntb=\@tempcntc
      \else\advance\@tempcntb\m@ne\@citeo
      \@tempcnta\@tempcntc\@tempcntb\@tempcntc\fi\fi}}\@citeo}{#1}}
\def\@citeo{\ifnum\@tempcnta>\@tempcntb\else\@citea\def\@citea{,}%
  \ifnum\@tempcnta=\@tempcntb\the\@tempcnta\else
  {\advance\@tempcnta\@ne\ifnum\@tempcnta=\@tempcntb \else\def\@citea{--}\fi
    \advance\@tempcnta\m@ne\the\@tempcnta\@citea\the\@tempcntb}\fi\fi}
\makeatother

\def\TeV{\ifmmode {\mathrm{\ Te\kern -0.1em V}}\else
                   \textrm{Te\kern -0.1em V}\fi}%
\def\GeV{\ifmmode {\mathrm{\ Ge\kern -0.1em V}}\else
                   \textrm{Ge\kern -0.1em V}\fi}%
\def\MeV{\ifmmode {\mathrm{\ Me\kern -0.1em V}}\else
                   \textrm{Me\kern -0.1em V}\fi}%
\def\keV{\ifmmode {\mathrm{\ ke\kern -0.1em V}}\else
                   \textrm{ke\kern -0.1em V}\fi}%
\def\eV{\ifmmode  {\mathrm{\ e\kern -0.1em V}}\else
                   \textrm{e\kern -0.1em V}\fi}%

\let\gev=\GeV

\title{ \LARGE\bf A precise determination of top quark electro-weak couplings at the ILC operating at $\roots=500\,\GeV$} 
\author[1]{M.S.~Amjad}
\author[2]{M.~Boronat}
\author[1]{T.~Frisson}
\author[2]{I.~Garc\'{\i}a{~}Garc\'{\i}a}
\author[1]{R.~P\"oschl\thanks{Corresponding author: poeschl@lal.in2p3.fr}}
\author[2]{E.~Ros}
\author[1,2]{F.~Richard}
\author[1]{J.~Rou\"en\'e}
\author[2]{P.~Ruiz Femenia}
\author[2]{M.~Vos}
\affil[1]{\footnotesize LAL, CNRS/IN2P3, Universit\'e Paris Sud, F-91898 Orsay CEDEX, FRANCE}
\affil[2] {\footnotesize IFIC, Universitat de Valencia CSIC, c/ Catedr\'atico Jos\'e Beltr\'an, 2  46980 Paterna, SPAIN}

\begin{document}


\maketitle
\thispagestyle{fancy}

\begin{abstract}
Top quark production in the process $e^+e^- \rightarrow t\bar{t}$ at a future linear electron positron collider with polarised beams is a powerful tool to determine indirectly the scale of new physics.  The presented study, based on a detailed simulation of the ILD detector concept, assumes a centre-of-mass energy of $\roots=500$\,GeV and a luminosity of $\mathcal{L}=500\,\invfb$ equality shared between the incoming beam polarisations of $P_{e^{-,+}} =\pm0.8,\mp0.3$. Events are selected in which the top pair decays semi-leptonically. The study comprises the cross sections, the forward-backward asymmetry and the slope of the helicity angle asymmetry.  The vector, axial vector and tensorial $CP$ conserving couplings are separately determined for the photon and the $Z^0$ component. 
The sensitivity to new physics  would be dramatically improved with respect 
to what is expected from LHC for electroweak couplings.  

\end{abstract}












\def\thefootnote{\fnsymbol{footnote}}
\setcounter{footnote}{0}
%

\section{Introduction}\label{sec:intro}

The main goal of current and future machines at the energy frontier is to understand the nature of electro-weak symmetry breaking.
This symmetry breaking can be generated by the existence of a new strong sector, inspired by QCD, that may manifest itself at energies of around 1\,TeV. In all realisations of the new strong sector, as for example Randall-Sundrum models~\cite{Randall:1999ee} or compositeness models~\cite{Pomarol:2008bh}, Standard Model fields would couple to the new sector  with a strength that is proportional to their mass. For this and other reasons, the $t$~quark is expected to be a window to any new physics at the TeV energy scale. New physics will modify the electro-weak $\ttbar X$ vertex described in the Standard Model by {\em V}ector and {\em A}xial vector couplings $V$ and $A$ to the vector bosons $X=\gamma, Z^0$.

Generally speaking, an $e^+e^-$ linear collider (LC) can measure $t$~quark electro-weak couplings at the \% level. In contrast to the situation at hadron colliders, the leading-order pair production process $\ee\to t \bar t$ goes directly through the $t\bar{t} Z^0$ and $t\bar{t} \gamma$ vertices.  There is no concurrent QCD production of $t$~quark pairs, which increases greatly the potential for a clean measurement. In the literature there a various ways to describe the current at the $t\bar{t} X$ vertex. Ref.~\cite{Juste:2006sv} uses: 
\begin{equation}\label{eq:snow}
\Gamma_\mu^{ttX}(k^2,\,q,\,\bar{q}) = ie \left\{
  \gamma_\mu \, \left(  \widetilde{F}_{1V}^X(k^2)
                      + \gamma_5\widetilde{F}_{1A}^X(k^2) \right)
+ \frac{(q-\bar{q})_\mu}{2m_t}
    \left(  \widetilde{F}_{2V}^X(k^2)
          + \gamma_5\widetilde F_{2A}^X(k^2) \right)
\right\} .
\end{equation}
with $k^2$ being the four momentum of the exchanged boson and $q$ and $\bar{q}$ the four vectors of the $t$ and $\bar{t}$~quark. Further $\gamma_\mu$ with
$\mu=0,..,3$ are the Dirac matrices describing vector currents and $\gamma_5=i\gamma_0\gamma_1\gamma_2\gamma_3$ is the Dirac matrix allowing to introduce
an axial vector current into the theory

Applying the Gordon identity to the vector and axial vector currents in Eq.~\ref{eq:snow} the parametrisation of the $t\bar{t} X$ vertex can be written as: 
\begin{equation}
\Gamma^{\ttbar X}_{\mu}(k^2, q, \bar{q}) = -ie\left\{\gamma_{\mu}\left( F_{1V}^X(k^2) +\gamma_5 F_{1A}^X(k^2) \right)   + \frac{\sigma_{\mu\nu}}{2m_t}(q+\bar{q})^{\mu}  \left(  iF_{2V}^X(k^2) +\gamma_5 F_{2A}^X(k^2)  \right) \right\},
\label{eq:vtxvtt}
\end{equation}
with $\sigma_{\mu\nu}=\frac{i}{2}\leftp \gamma_{\mu}\gamma_{\nu} - \gamma_{\nu}\gamma_{\mu} \rightp$.
The couplings or form factors $\widetilde F^X_{i}$ and $F^X_{i}$ appearing in Eqs.~\ref{eq:snow} and~\ref{eq:vtxvtt} are related via 
\begin{equation}
\label{eq:rel1}
\widetilde F^X_{1V} = -\left( F^X_{1V}+F^X_{2V} \right) \, , \qquad
\widetilde F^X_{2V}  =  F^X_{2V} \, , \qquad
\widetilde F^X_{1A} = -F^X_{1A} \, , \qquad
\widetilde F^X_{2A} =  -iF^X_{2A} \, .
\end{equation}

Within the Standard Model the $F_i$ have the following values:
\begin{equation}
F_{1V}^{\gamma,SM}=-\frac{2}{3},\,F_{1A}^{\gamma,SM}=0,\,F_{1V}^{Z,SM}=-\frac{1}{4s_wc_w}\left(1-\frac{8}{3}s^2_w \right),\,F_{
1A}^{Z,SM}=\frac{1}{4s_wc_w},
\label{eq:ffactors}
\end{equation}
with $s_w$ and $c_w$ being the sine and the cosine of the Weinberg angle $\theta_W$. 

All the expressions above are given at Born level. Throughout the article no attempt will be made to go beyond that level.
The coupling $F^{\gamma}_{2V}$ is related via $F^{\gamma}_{2V}=Q_{t}(g-2)/2$ to the anomalous magnetic moment $(g-2)$ with $Q_t$ being
the electrical charge of the $t$~quark. The coupling $F_{2A}$ is related to the dipole moment $d=(e/2mt)F_{2A}(0)$ that violates the combined {\em C}harge and {\em P}arity symmetry $CP$. Due to the $\gamma/Z^0$ interference the $t$ pair production is sensitive to the sign of the individual form factors. This is in contrast to couplings of $t$~quarks to $\gamma$ and $Z^0$ individually where the form factors enter only quadratically.  The authors of~\cite{Baur:2004uw} use LEP data to set indirect constraints on the electric form factors $F_{1V,A}^Z$. These limits are sufficient to exclude an ambiguity for $F_{1A}^Z$ but not for $F_{1V}^Z$. In~\cite{Baur:2004uw} also rough estimates for $|F^{\gamma,Z}_{2V}|$ are given. 

Today, the most advanced proposal for a linear collider is the International Linear Collider, ILC~\cite{bib:ilc-tdr-dbd}, which can operate at centre-of-mass energies between the $Z^0$~pole to 1\,TeV. The ILC provides an ideal environment to measure these couplings. The $\ttbar$ pairs would be copiously produced, several 100,000 events at $\roots=500\,\GeV$ for an integrated luminosity of $500\,\invfb$. It is possible to almost entirely eliminate the background from other Standard Model processes. The ILC will allow for polarised electron and positron beams.  With the use of polarised beams, $t$ and $\bar t$ quarks oriented toward different angular regions in the detector are enriched in left-handed or right-handed $t$~quark helicity~\cite{Parke:1996pr}.  This means that the experiments can independently access the couplings of left- and right-handed chiral parts of the $t$~quark wave function to the $Z^0$ boson and the photon.  In principle, the measurement of the cross section and forward-backward
asymmetry $\afbt$ for two different polarisation settings allows extracting both, the photon and $Z^0$ couplings of the $t$~quark for each helicity state. This study introduces the angle of the decay lepton in semi-leptonic decays of the $\ttbar$ in the rest frame of the $t$~quark. This angle will be called the {\em helicity angle}. The slope of the resulting angular distribution is a measure for the fraction of $t$~quarks in left-handed helicity state, $t_L$ and right-handed helicity state, $t_R$, in a given sample. There are therefore six independent observables
\begin{itemize}
\item The cross section;
\item The forward backward asymmetry $\afbt$;
\item The slope of the distribution of the helicity angle; 
\end{itemize}
for two beam polarisations. For the extraction of the six $CP$ conserving form factors defined for the $Z^0$ and the photon, $F_{1V},\,F_{1A}$ and $F_{2V}$, the following observations have to be taken into account: Close to the $\ttbar$ threshold the observables depend always on the sum $F_{1V}+F_{2V}$. Therefore a full disentangling of the form factors will be imprecise for energies below about $1\,\TeV$. Hence, in the present study either the four form factors $F_{1V,A}$ are varied simultaneously, while the two $F_{2V}$ are kept at their Standard Model values or vice versa. Throughout this article the $CP$ violating form factors $F^{X}_{2A}$ will be kept at their Standard Model values.

This article is organised as follows. After this introduction the relations between the observables and the form factors are outlined before the experimental environment and the used data samples will be introduced. After that the selection of semi-leptonic decays of the $\ttbar$ pair will be presented and the selection efficiencies will be given. The determination of $\afbt$ will be followed by the extraction of the slope of the distribution of the helicity angle. Potential systematic effects from experiment and theory uncertainties will be outlined. Finally the six $CP$ conserving form factors will be extracted as explained above. This study goes therefore beyond earlier studies published in~\cite{bib:Abe:2001nq,AguilarSaavedra:2001rg}, 

\section{Top quark production at the ILC}

\begin{figure}[h!]
\begin{center}
    \mbox{
      \begin{tabular}{cc}
        \subfigure[$t\bar{t}$ pair production]{\label{fig:diagrams_a} \includegraphics[width=0.3\columnwidth]{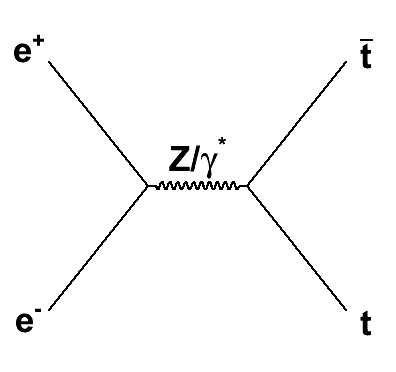}}
        \subfigure[Single top production]{\label{fig:diagrams_c} \includegraphics[width=0.3\columnwidth]{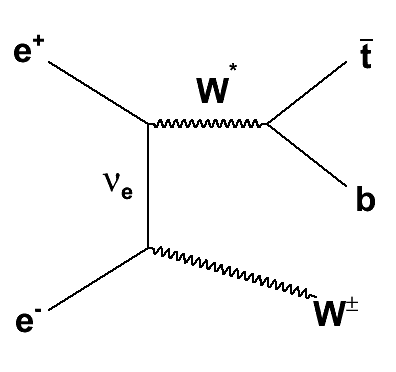}}
      \end{tabular}
    }
\end{center}
\caption{Four diagrams that contribute to the $e^+ e^- \rightarrow l \nu b \bar{b} q' \bar{q}$ production: (a) Born-level $t \bar{t}$ pair production, (b) box diagram higher-order contribution to the same process (c) single top production (d) triple gauge boson production. }
\label{fig:feyn_diagrams}
\end{figure}

The dominant source of top quark production at a 500~\gev{} $e^+e^-$ collider
is electro-weak production. The Born-level diagram is presented in Figure~\ref{fig:diagrams_a}. 


The decay of the top quarks proceeds predominantly through  $t \rightarrow W^{\pm} b$. The subsequent decays of the $W^{\pm}$ bosons to a charged lepton and a neutrino or a quark-anti-quark pair lead to a  six-fermion final state. The study presented in this article focuses on the 'lepton+jets' final state $ l^{\pm} \nu b \bar{b} q' \bar{q}$.

Several other Standard Model processes give rise to the same final state. The most important source is single-top production through the process
$e^+e^- \rightarrow W W^* \rightarrow W t \bar{b} \rightarrow l^{\pm} \nu b \bar{b} q' \bar{q}$. One of the diagrams contributing to this process is
presented in Figure~\ref{fig:diagrams_c}. Another relevant source is the $Z^0 WW$ production. Due to the coupling of initial state electrons or positrons to to $W$ bosons both sources contribute nearly exclusively in a configuration with left handed polarised electron beams and right handed polarised positron beams. 

In that case single-top and $Z^0 WW$ production would yield a production rate of order 10\% and 5\%, respectively, of that of the pair production diagram of 
Figure~\ref{fig:diagrams_a}. Experimentally, $Z^0 WW$ production can be distinguished rather efficiently from top quark pair production. The separation
between single-top production and top quark pair production is much more difficult. At a more fundamental level, interference between these 
processes renders the separation physically meaningless. A realistic experimental strategy must therefore consider the six-fermion production
process $e^+e^- \rightarrow l^{\pm} \nu b \bar{b} q' \bar{q}$.

\subsection{Observables and Form Factors}

According to~\cite{MoortgatPick:2005cw}, the cross section for any process in $\epem$ collisions in case of polarised beams can be written as
\be
\sigma_{\pem,\pep} = \frac{1}{4}\left[(1- \pem \pep)(\sigma_{-,+}+\sigma_{+,-})+(\pem-\pep)(\sigma_{+,-}-\sigma_{-,+})\right] 
\label{eq:tot-cross}
\eeqn

In this equation the symbols $-$ and $+$ indicate full polarisation of the incoming beams with electrons and positrons of left-handed, $L$, or right-handed, $R$, helicity, respectively. The configurations $\sigma_{-,-}$ and $\sigma_{+,+}$ have been neglected due to helicity conservation at the electron vertex in the high energy limit. The degree of polarisation of the incoming beams is expressed by $\pem$, for electrons, and $\pep$, for positrons.

In case of polarised beams Ref.~\cite{Schmidt:1995mr} suggests to express the form factors introduced in Sec.~\ref{sec:intro} in terms of the helicity of the incoming electrons,  
\begin{eqnarray}
{\cal F}^L_{ij}&=& -F^\gamma_{ij}+
\Bigl({-{1\over2}+s_w^2\over s_wc_w}\Bigr)\Bigl({s\over
s-m_Z^2}\Bigr)F^Z_{ij}\nonumber\\
{\cal F}^R_{ij} &=&-F^\gamma_{ij}+\Bigl({s_w^2\over s_wc_w}\Bigr)
\Bigl({s\over s-m_Z^2}\Bigr)
F^Z_{ij}\ ,\label{combffs}
\end{eqnarray}
with $i=1,2$ and $j=V,A$ and $m_Z$ being the mass of the $Z^0$~boson.
The Born cross section for $\ttbar$~quark production for electron beam polarisation $I=L,R$ reads 
\beq
\sigma_{I}=2{\cal A}N_c\beta\left[  (1+0.5\gamma^{-2})(\fonevI)^2  + (\foneaI)^2+3\fonevI \ftwovI \right],
\eeqn
where ${\cal A}=\frac{4\pi\alpha^2}{3s}$ with the running electromagnetic coupling $\alpha(s)$ and $N_c$ is the number of quark colours. Furthermore $\gamma$ and $\beta$ are the Lorentz factor and the velocity, respectively. The term $\foneaI = \beta {\cal F}^{I}_{1A}$ describes the reduced sensitivity to axial vector couplings near the $\ttbar$ production threshold. The cross sections at the Born level of the signal process $\epem \rightarrow \ttbar$ and the main Standard Model background processes at a centre-of-mass energy of $500\,\GeV$ are summarised in Table~\ref{tab:procs}. 
\begin{table}[!h]
\begin{center}
\centerline{\begin{tabular}{|c|c|c|c|c|}
\hline
Channel & $\sigma_{unpol.}$ [fb]& $\sigma_{-,+}$ [fb]& $\sigma_{+,-}$ [fb] & $\mathrm{A^{SM}_{LR}}$\%\\ 
\hline
$\ttbar$ & 572 & 1564 & 724 & 36.7 \\
\hline
$\mu \mu$ & 456 & 969 & 854 & 6.3 \\
\hline
$\sum_{\mathrm{q=u,d,s,c}} \qq$ & 2208 & 6032 & 2793 & 36.7 \\
\hline
$b \bar{b}$ & 372 & 1212 & 276 & 62.9 \\
\hline
$\gamma \Zzero$ & 11185 & 25500 & 19126 & 14.2 \\
\hline
$W W$ & 6603 & 26000 & 150 & 98.8\\
\hline
$\Zzero \Zzero$ & 422 & 1106 & 582 & 31.0\\
\hline 
$\Zzero W W$ & 40 & 151 & 8.7 & 89\\
\hline
$\Zzero \Zzero \Zzero$ & 1.1 & 3.2 & 1.22 & 45\\
\hline
\end{tabular}}
\end{center}
\caption{\sl Unpolarised cross-sections and cross-sections at the Born level for 100\% beam polarisation for signal and background processes. The last column gives the left right asymmetry as expected from the Standard Model.}
\label{tab:procs}
\end{table}%

The forward-backward asymmetry $\afbt$ can be expressed as 
\beq
(\afbt)_{I} =\\ \frac{-3 \foneaI (\fonevI + \ftwovI )}{  2\left[  (1+0.5\gamma^{-2})(\fonevI)^2  + (\foneaI)^2+3\fonevI \ftwovI \right]    },
\eeqn
which in the Standard Model takes the values $(\afbt)_{L} =0.38 $ and  $(\afbt)_{R} =0.47$.

The fraction of right-handed tops is given by the following expression:
\begin{footnotesize}
\beq
(F_R)_{I} =  \frac{ (\fonevI)^2(  1+0.5\gamma^{-2} )  + (\foneaI)^2 + 2\fonevI \foneaI + \ftwovI (3\fonevI + 2 \foneaI) -\beta\fonevI\mathfrak{Re} (\ftwoaI)} {  2\left[  (1+0.5\gamma^{-2})(\fonevI)^2  + (\foneaI)^2+3\fonevI \ftwovI \right] }.
\label{eq:fri}
\eeqn
\end{footnotesize}
The values expected in the Standard Model are $(F_R)_{L}=0.25$ and $(F_R)_{R}=0.76$. 
The Eq.~\ref{eq:fri} contains a $CP$ violating term proportional to $\mathfrak{Re} (\ftwoaI)$. This term will not be determined in the present study but can also be precisely estimated using $CP$ violating observables, see later in Tab.~\ref{tab:tab2}. This implies that $CP$ conserving form factors can be fully disentangled without the assumption of $CP$ conservation.

\section{Theory uncertainties}
\label{sec:theory}

The extraction of form factors requires precise predictions of the 
inclusive top quark pair production rate and of several differential 
distributions. In this section the state-of-the-art
calculations and estimate theoretical uncertainties are briefly reviewed.

\begin{figure}[h]
  \begin{center}
    \mbox{
      \begin{tabular}{cc}
        \subfigure[Perturbation series]{\label{fig:qcd_uncertainty_a}\includegraphics[width=0.49\columnwidth]{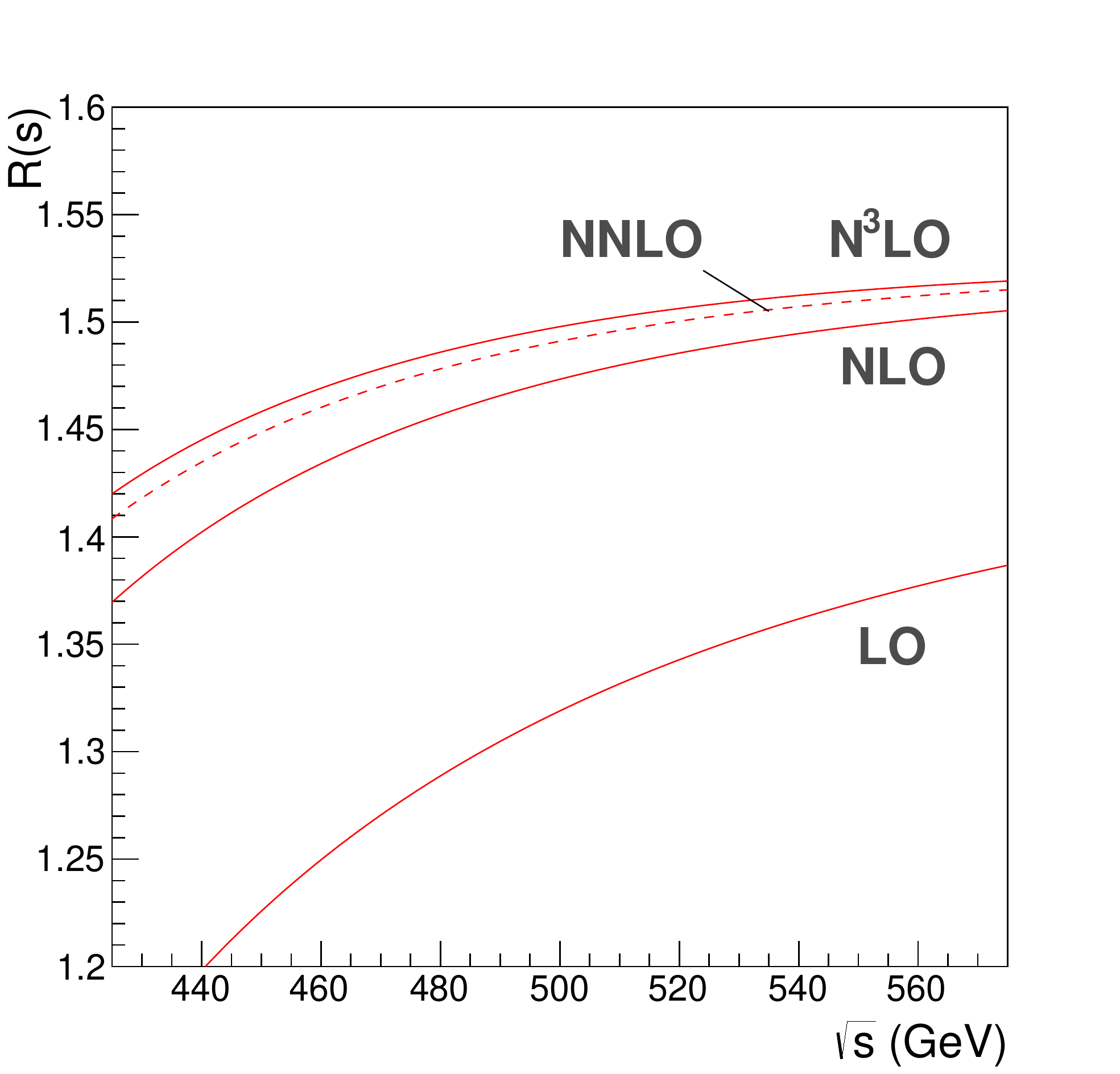}}
        \subfigure[Scale variations]{\label{fig:qcd_uncertainty_b}\includegraphics[width=0.49\columnwidth]{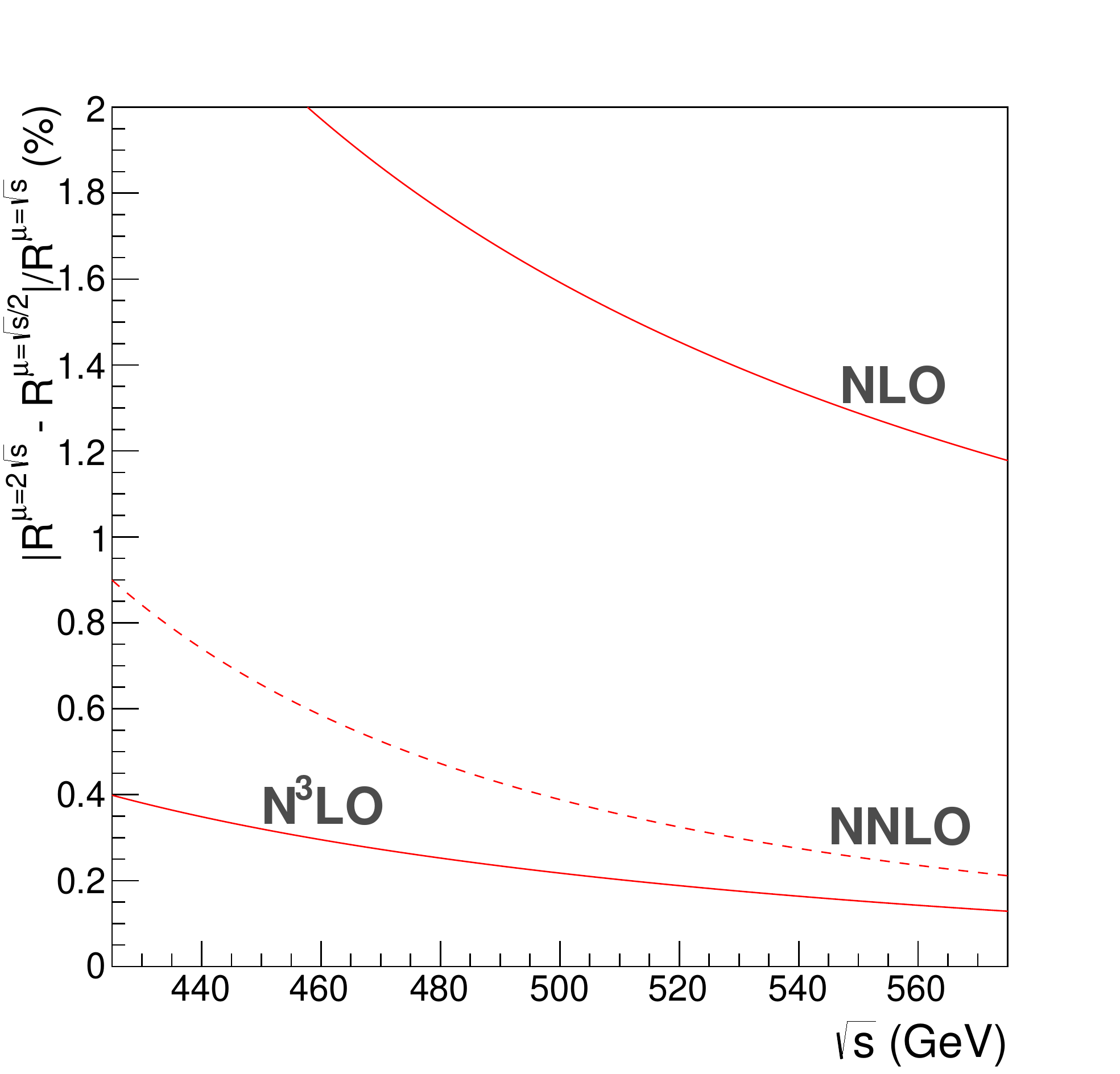}}
      \end{tabular}
    }
  \end{center}
\caption{(a) The prediction for the ratio $R(s)$ of the $t \bar{t}$ 
production rate and that of massless fermions, as a function of 
centre-of-mass energy $\sqrt{s}$. (b) 
The maximum variation (in \%) in the cross section prediction due to 
alternative choices for the renormalisation scale: $\mu = 2 \sqrt{s}$. 
These figures present a compilation of results reported in 
References~\cite{Kiyo:2009gb,Hoang:2008qy,Bernreuther:2006vp}.}
\label{fig:diagrams}
\end{figure}

The QCD corrections to $e^+e^- \rightarrow t \bar{t}$ production are
known up to $N^3LO$ for the inclusive cross section~\cite{Kiyo:2009gb}, 
and to $NNLO$ for the forward backward asymmetry 
$A_{FB}$~\cite{Bernreuther:2006vp}. The perturbative series shows good 
convergence. In Figure~\ref{fig:qcd_uncertainty_a} the $LO$, $NLO$, $NNLO$ 
and $N^3LO$ predictions for the ratio $R(s)$ of the total cross section
for $e^+e^- \rightarrow t \bar{t}$ to that for massless fermion pair production
are shown in the centre-of-mass energy range around 500~\gev{}. 

The $N^3LO$ correction to the total cross-section is below 1 \%. An estimate 
of the size of the next order - obtained from the conventional variation of 
the renormalisation scale by a factor two and one half - yields 0.3 \%. It can therefore be
concluded that the uncertainty of today's state-of-the-art 
calculations is at the per mil level.

In a similar manner the QCD corrections to the prediction of
differential distributions and quantities such as the forward-backward 
asymmetry can be estimated. The size of the $N^3LO$ correction to $A_{FB}$ is estimated using the 
scale variation to be smaller than 1\%, see also the discussion in e.g.~\cite{Bernreuther:2006vp}.

Electro-weak (EW) corrections to the same process have also been calculated. 
A full one-loop calculation is presented in Reference~\cite{Fleischer:2003kk}.
The correction to the total cross section is found to be approximately 5\%.
The EW correction to the forward-backward asymmetry is large, approximately
10\%~\cite{Fleischer:2003kk,Khiem:2012bp}. 

The above discussion refers to corrections to the 
process $e^+e^- \rightarrow t \bar{t}$. Further corrections 
of order $\Gamma_t/m_t \sim $ 1\% are expected to appear if the decay of the top
quarks is included in the calculation.

It can be concluded that the state-of-the-art calculations of QCD corrections offer 
the precision required for this study. Uncertainties
are under relatively good control, with uncertainties to the cross section
of the order of a few per mil and order 1\% on the forward-backward asymmetry. Electro-weak (one-loop) corrections are large.
Further work is needed to estimate the size of the two-loop correction 
and, ultimately, to calculate this contribution. Currently these aspects are discussed with theory groups.

\section{Experimental environment and data samples}
The International Linear Collider is a project for a linear electron-positron accelerator reaching a centre-of-mass energy of up to 1\,TeV.  For a detailed description of the machine the reader is referred to~\cite{bib:ilc-tdr-dbd}. 
For the studies presented in this article it is important
to emphasise that the machine can deliver polarised electron and positron 
beams. At a centre-of-mass energy of $\roots=500\,\GeV$ the envisaged degree 
of polarisation is 80\% in case of electrons and 30\% in case of positrons.

The ILD detector is designed as a detector for Particle Flow. This means that the jet energy measurement is based on the measurement of individual particles~\cite{Brient:2002gh}.
A detailed description of the current model of the ILD detector can be found elsewhere~\cite{bib:ilc-tdr-dbd}.
The $z$-axis of the right handed co-ordinate system is given by the direction of the incoming electron beam. Polar angles given in this note are defined with respect to this axis.
The most important sub-detectors for this study  are described in the following.
\begin{itemize}
\item The vertex detector consists of three double layers of silicon extending between 16\,mm and 60\,mm in radius and between 62.5\,mm and 125\,mm in $z$ direction. It is designed for an impact parameter resolution of $\sigma_{r\phi}=\sigma_{rz}=5\oplus 10/(p {\rm sin}^{\frac{3}{2}}\theta)\,\mathrm{\mu m}$.
\item The measurement of charged tracks is supported by an inner Silicon Tracker (SIT) in the central region and by a set of silicon disks in forward 
direction, i.e. towards large absolute values of $cos\theta$.
\item The ILD detector contains a large Time Projection Chamber (TPC) with an inner sensitive radius of 395\,mm and an outer sensitive radius of 1743\,mm.  The half length in $z$ is 2250\,mm. Recent simulation studies confirm that the momentum of charged particle tracks can be measured to a precision of $\delta(1/P_T)\sim 2\times10^{-5}\,\GeV^{-1}$. Here $P_T$ denotes the transverse component of the three  momentum $P$ of the particles.
\item The electromagnetic calorimeter is a SiW sampling calorimeter. Its longitudinal depths of 24\,${\rm X_{0}}$ allows
for the complete absorption of photons with energies of up to 50 GeV as relevant for the studies here. The simulated energy
resolution of the electromagnetic calorimeter is  $\frac{\Delta E}{E}=15\%/\sqrt{E\,\mathrm{[GeV]}}$
\item The hadronic calorimeter surrounds the electromagnetic calorimeter and comprises 4.5 interaction length $\lambda_I$.  

Two proposals exist for the hadronic calorimeter. A semi-digital variant consisting of steel absorbers and gas RPC chambers with a pixel size of $1 \times 1$\,${\rm cm^2}$ as active material. The second one features scintillating tiles with size of $3 \times 3$\,${\rm cm^2}$ as active material. The latter option is employed in the present work.

\end{itemize}

Simulation studies show that the high precision tracking together with the highly granular calorimeters allow for achieving a jet energy resolution of 3-4\% for jet energies up to 1.5\,TeV~\cite{bib:ilc-tdr-dbd,Linssen:2012hp}.

\subsection{Event generation and technical remarks}


Signal and background events are generated with version 1.95 of the {\tt WHIZARD} event  generator~\cite{Kilian:2007gr,Moretti:2001zz} in the form of six for the signal, or two and four fermion final states for Standard Model background. The most relevant background contributions are the $WW$ and $b\bar{b}$ final states as well as fully hadronic and fully leptonic decays of $\ttbar$~pairs. The generated events are then passed to the PYTHIA simulation program to generate parton showers and subsequent hadronisation. In case of the signal sample events are flagged for which the difference between the invariant masses of the three fermion systems forming a top from WHIZARD and the input $t$ mass to WHIZARD of $174\,\GeV$ is smaller than $5\Gamma_t$. Here $\Gamma_t$ is the total decay width of the $t$~quark. By this only about 70\% of the events generated by WHIZARD are recognised as $\ttbar$ events and treated accordingly by PYTHIA. The different hadronisation of genuine $\ttbar$ events may introduce a systematic uncertainty, which will have to be estimated at a later stage. It is however expected to be reasonable small.



The study has been carried out on a fully polarised sample.  Realistic values of the beam polarisations at the ILC at $\roots=500\,\GeV$ are however  $\pem, \pep =\pm0.8,\mp0.3$. 
The cross section and therefore its uncertainty scales with the polarisation according to Eq.~\ref{eq:tot-cross}.  The observables $\afbt$ and $\lhel$ vary only very mildly with the beam polarisation. Again, the reduced cross section leads to a higher statistical error for non-fully polarised beams. This will be correctly taken into account in the uncertainty of the results. 

Events corresponding to a luminosity of $250\,\invfb$ for each of the polarisation configurations were subject to a full simulation of the ILD detector and subsequent event reconstruction using the version {\tt ILD\_o1\_v05} of the ILC software. 


\section{Event selection}\label{sec:sel}



The analysis starts out from the studies presented in detail in~\cite{ild:bib:benchmark:doublet}.  
The produced $t$($\bar{t}$)-quark decays almost exclusively in to a $\bottom W$ pair. The $b$~quark hadronises giving rise to a jet. The $W$~boson can decay {\em hadronically} into light quarks, which turn into jets, or {\em leptonically} into a pair composed by a charged lepton and a neutrino. The {\em semi-leptonic process} is defined by events in which one $\Wboson$ decays hadronically while the other one decays leptonically, i.e.
 \begin{equation}
\ttbar \rightarrow (\bottom \Wboson) (\bottom \Wboson) \rightarrow (\bottom \quark \quark') (\bottom \ell \nu) 
\end{equation}
In the Standard Model  the fraction of semi-leptonic final states in  $\epem \rightarrow \ttbar$ is about 43\%. 

\begin{figure}[h]
  {\centering 
 \subfigure[Durham]{\label{fig:jetrecowhad_a} \includegraphics[width=0.45\textwidth]{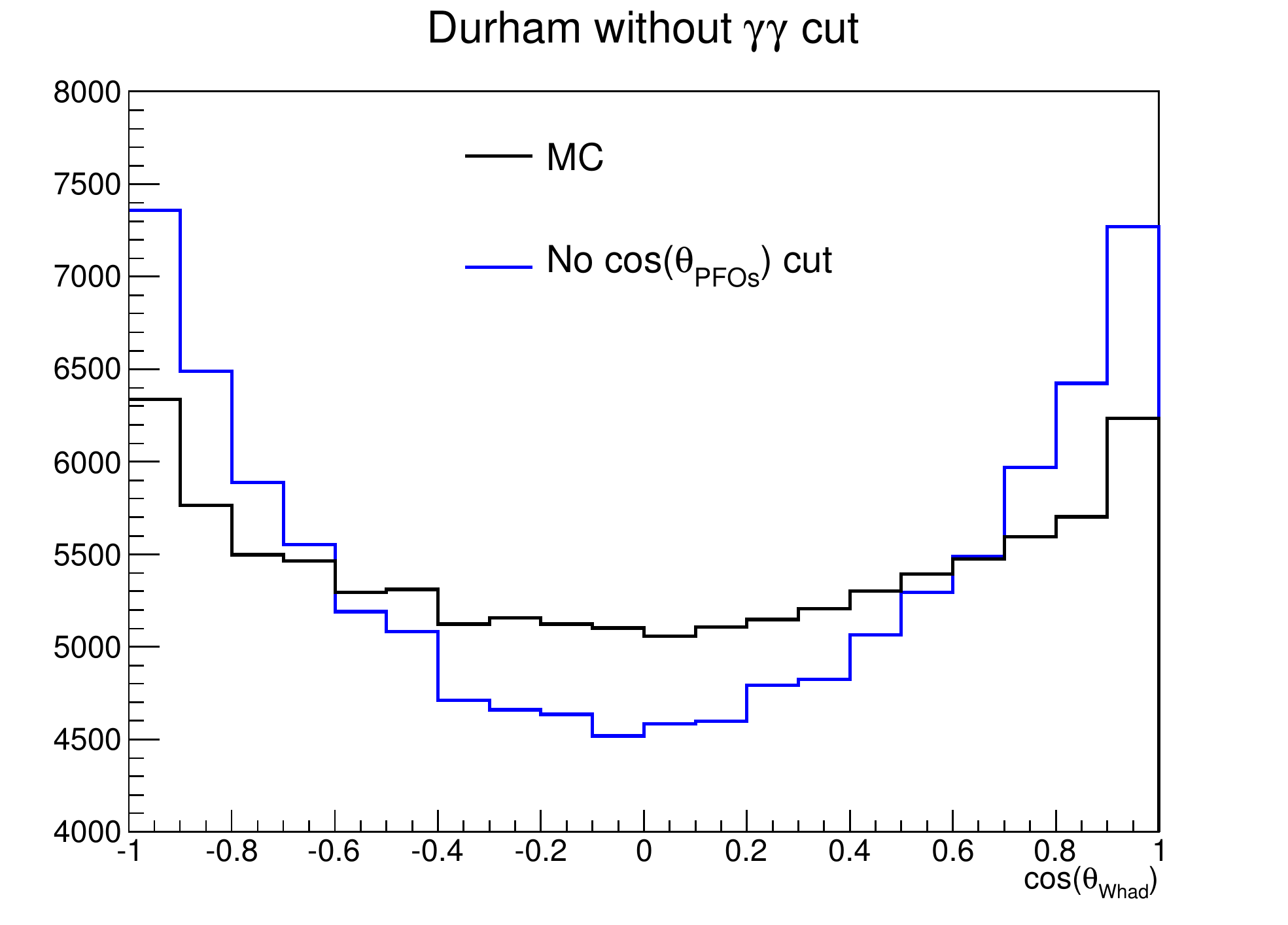}}
\subfigure[$k_t$, $R=1.5$]{\label{fig:jetrecowhad_d} \includegraphics[width=0.45\textwidth]{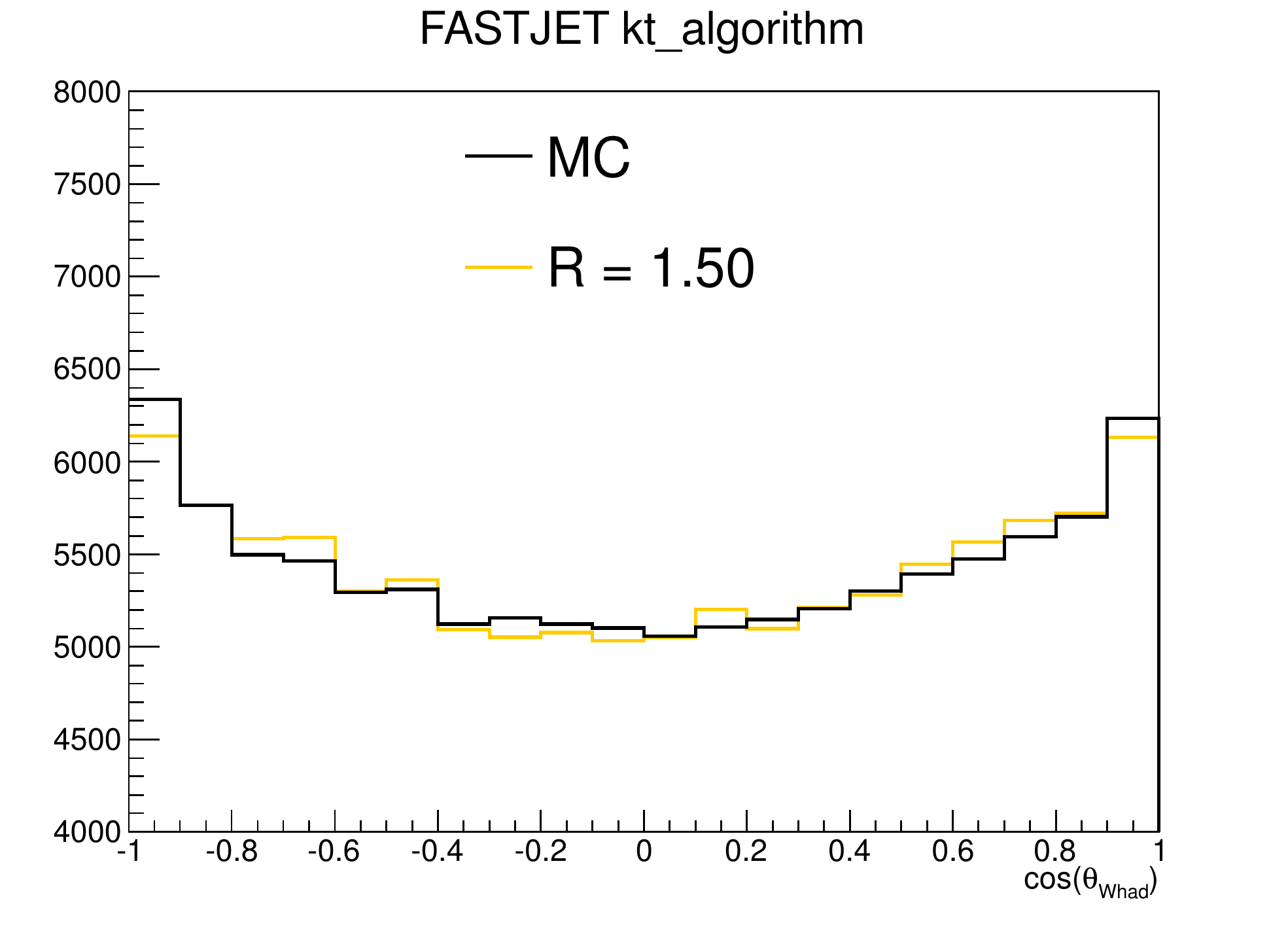}} \caption{The polar angle distribution of the hadronically decaying $W$ for four different jet algorithms.}
        \label{fig:jetrecowhad}
        }
        \end{figure}

A number of processes known collectively as multi-peripheral $\gamma \gamma \rightarrow ${\it hadrons} production yield a small number of additional particles (typically 1.7 low-multiplicity events per bunch crossing for the ILC).  The polar angle distribution of these particles is markedly forward.
Particles from $\gamma \gamma \rightarrow ${\it hadrons} tend to be harder and can reach the outer layers of the detector and affect the overall detector performance, in particular jet reconstruction. These particles are similar to beam remnants as present in hadron collisions. It is therefore intuitive to employ a jet algorithm used in hadron collisions that separate the beam jet from the hard interaction. In the present study the best results are obtained with the the longitudinally invariant $k_t$ algorithm~\cite{Catani:1993hr,Ellis:1993tq}. 
This is demonstrated in Fig.~\ref{fig:jetrecowhad}. 
The figure shows the reconstructed polar angle distribution of the hadronically decaying $W$ boson from $\\tbar$ pairs compared with the generated distribution. The result is shown for the ''traditional'' Durham~\cite{Stirling:1991ds} algorithm and for the longitudinally invariant $k_t$ algorithm with a jet radius of $R=1.5$. The improvement achieved by the longitudinally invariant $k_t$ algorithm is obvious.  After the removal of the $\gamma \gamma \rightarrow ${\it hadrons} background the event is processed further with the Durham algorithm. Further beam induced background such as electron-positron pairs have not been studied in the present article but a detailed study presented in~\cite{Vogel:1181870} demonstrates that the induced number of background hits in the vertex detector and the TPC as well as the related neutron fluxes are uncritical for the detector performance. 


The charged lepton allows for the determination of the  $t$~quark charge. The $t$~quark mass is reconstructed from the hadronically decaying $\Wboson$ which is combined with one of the $\bottom$-quark jets.
In general leptons are identified using typical selection criteria. The lepton from the $W$~boson decay is either the most energetic particle in a jet  or has a sizable transverse momentum w.r.t. neighboured jets. More specific the following criteria are applied
\be
x_T=p_{T,lepton}/M_{jet} > 0.25\,\,\,\,\,\mathrm{and}\,\,\,\, z=E_{lepton}/E_{jet} >0.6,
\eeqn
where $E_{lepton}$ is the energy and $p_{T,lepton}$ the transverse momentum of the lepton within a jet with energy $E_{jet}$ and mass $M_{jet}$.
The decay lepton in case of $e$ and $\mu$ can be identified with an efficiency of about 85\%, where the selection has a tendency to reject low momentum leptons. The $\tau$~leptons can decay themselves into $e$ or $\mu$, which are collinear with the produced $\tau$ but have lower momentum than primary decay leptons. Taking into account the $\tau$~leptons, the efficiency to identify the decay lepton is about 70\%.

\begin{figure}[ht]
\begin{center}
\includegraphics[width=\textwidth]{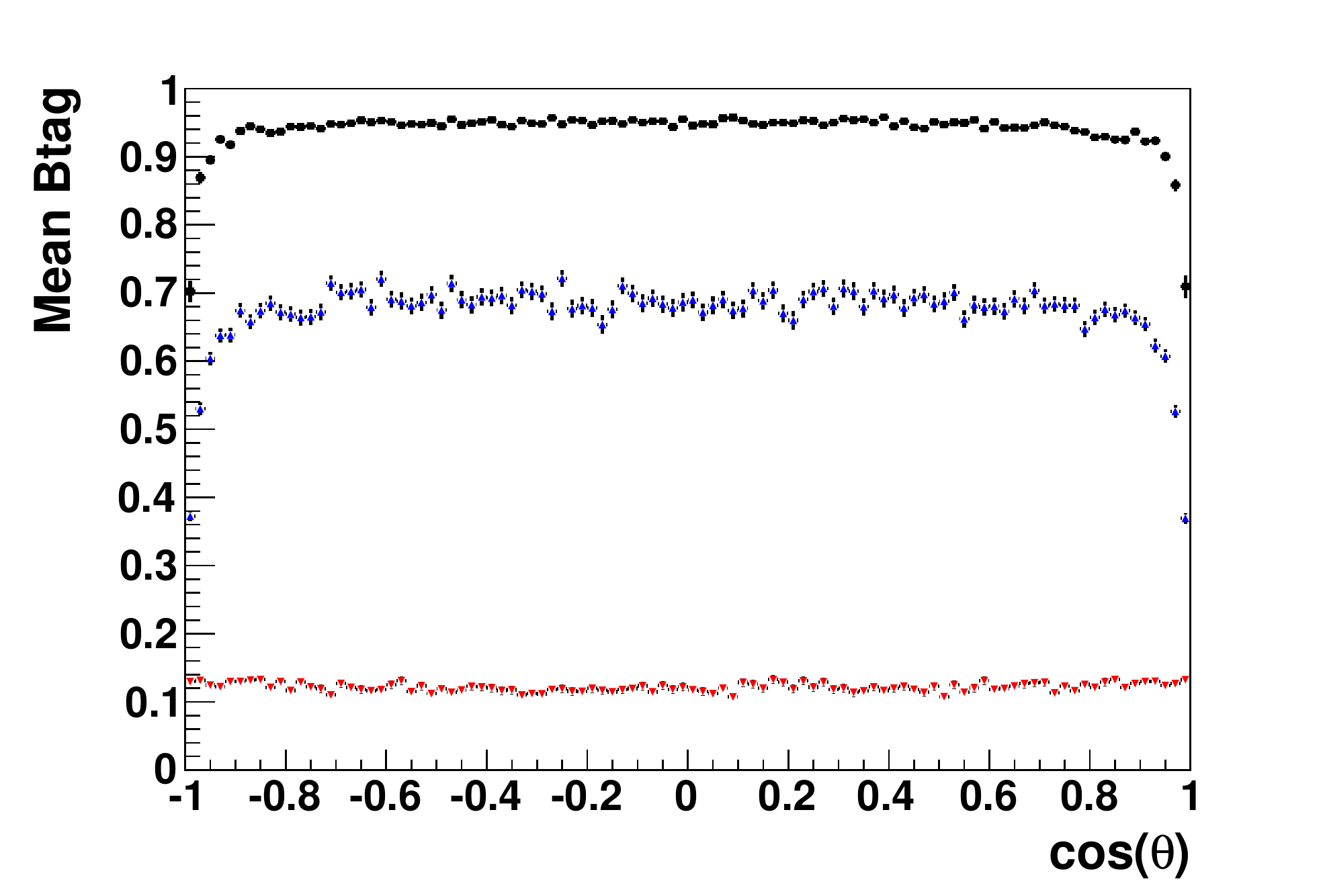}
\caption{\sl The $b$-tag values as a function of the polar angle of the jets. The two highest $b$-tag values (black and blue dots) are associated to $b$ quark jets. The third set of values (red dots) is obtained for jets from light quarks.}
\label{fig:bt}
\end{center}
\end{figure}

The identified lepton is removed from the list of reconstructed particles and the remaining final state is again clustered into four jets. Two of these must be identified as being produced by the $\bottom$-quarks of the $\tpq$~quark decay. 
The $b$-likeness or {\em b-tag} is determined with the {\tt LCFIPlus} package, which uses information of the tracking system as input. Secondary vertices in the event are analysed by means of the jet mass, the decay length and the particle multiplicity. 
The jets with the highest $b$-tag values are selected. As shown in Fig.~\ref{fig:bt} the higher $b$-tag value is typically 0.92 while the smaller one is still around 0.65. Both values are clearly distinct from those obtained for jets from light quarks. Their $b$-tag value is around 0.14.

\begin{figure}[h]
        {\centering
        \subfigure[\small Momentum of $b$~jet at top rest frame.]{\includegraphics[width=0.49\textwidth]{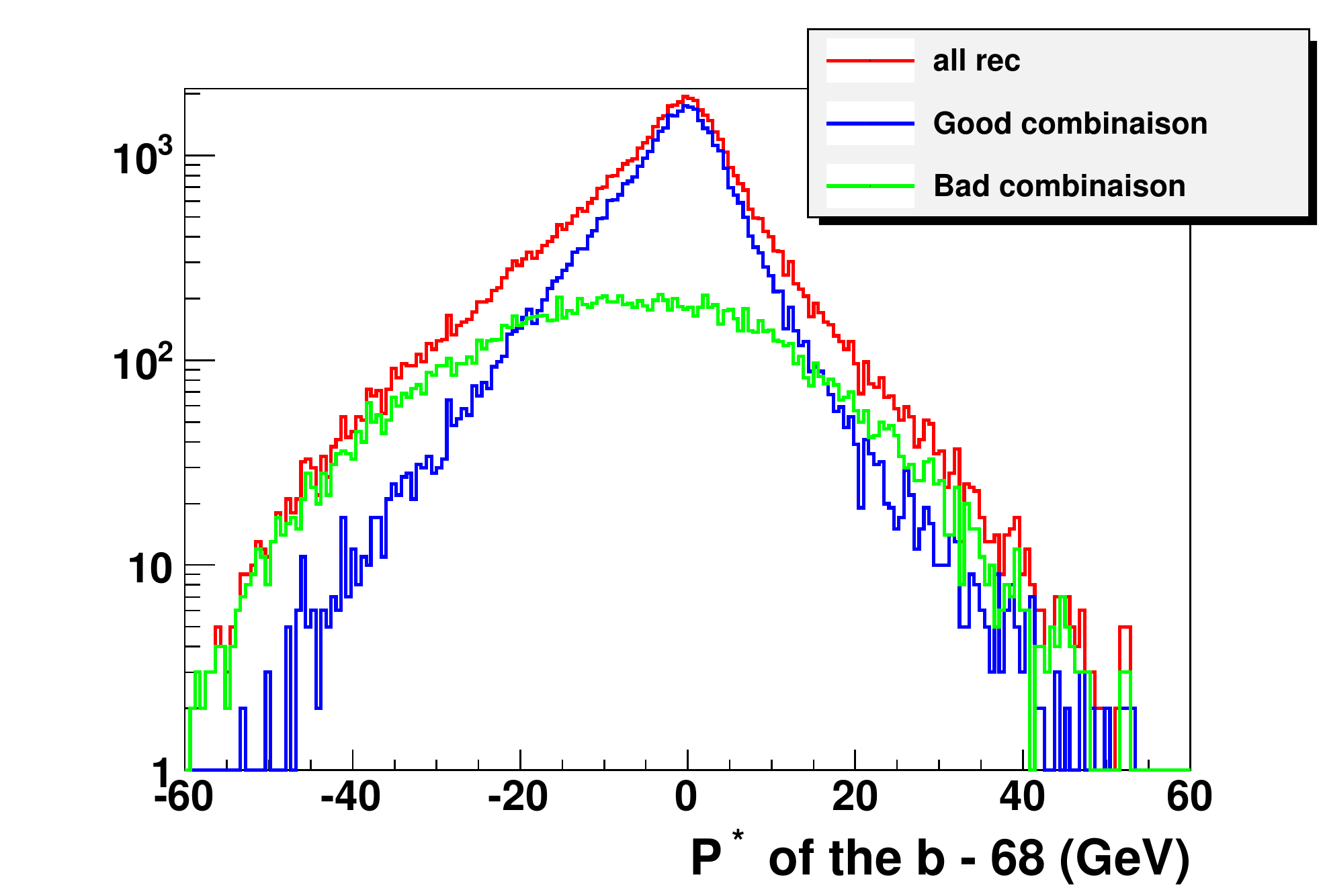}}
        \subfigure[\small Angle between b-jet and W.]{\includegraphics[width=0.49\textwidth]{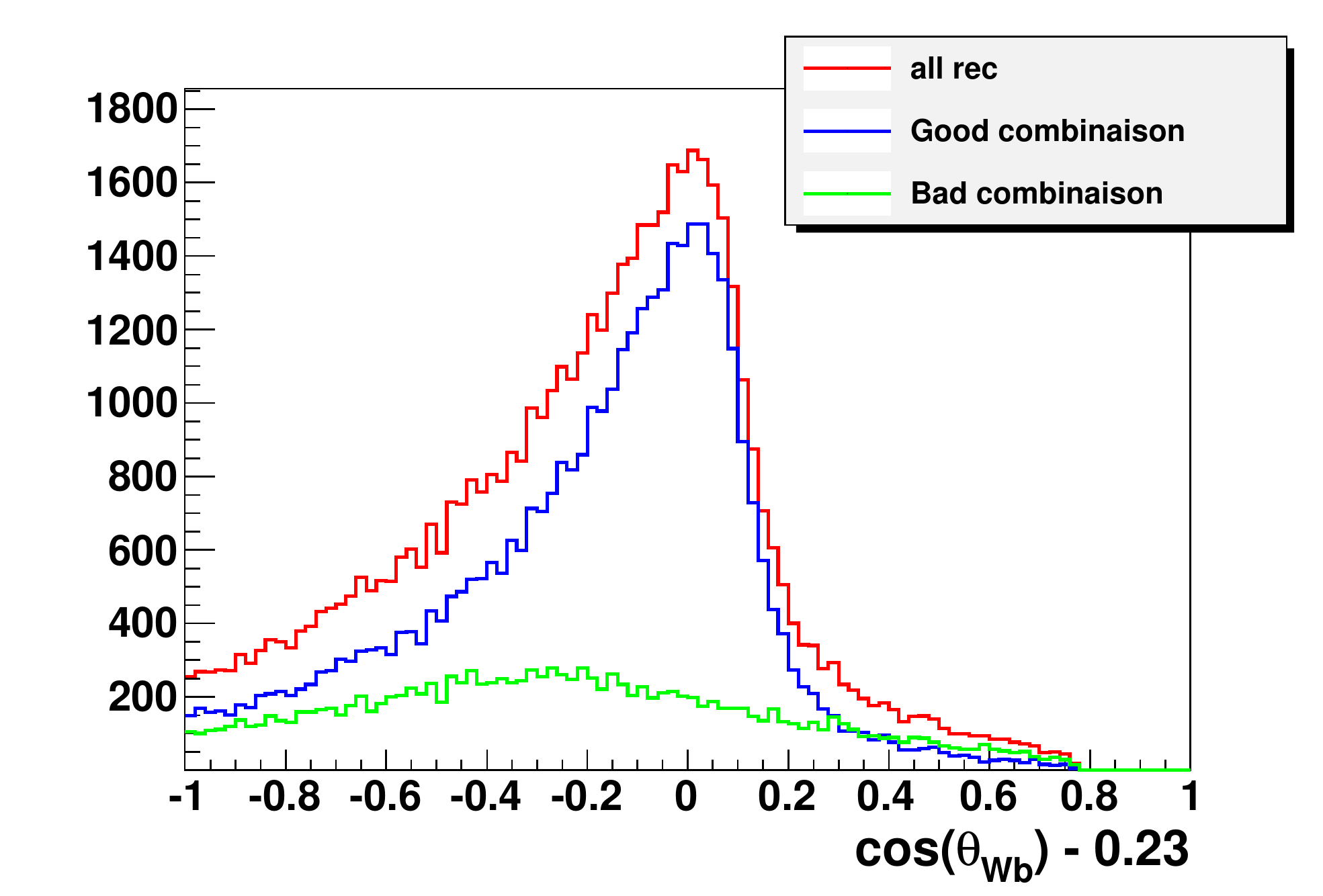}}
        \caption{\sl Distributions of the momentum of the $b$~quark jet in the centre-of-mass frame of the $t$ quark, $p_{b}^{\ast}$ and the cosine of the angle $\theta_{bW}$ between the $b$~quark and the $W$~boson.} 
        \label{fig:pstarwb}}
\end{figure}

These values are nearly independent of the polar angle of the $b$~quark jet but drop towards the acceptance limits of the detector. Finally, the two remaining jets are associated with the decay products of the $W$~boson. The signal is reconstructed by choosing that combination of $b$~quark jet and $W$~boson that minimises the following equation:
\begin{equation}
d^2 = \left( \frac{m_{cand.}-m_{\tpq}}{\sigma_{m_{\tpq}}} \right)^2 + \left( \frac{E_{cand.} - E_{beam}}{\sigma_{E_{cand.}}}\right )^2 + \left(\frac{p_b^{\ast} -68}{\sigma_{p_{b}^{\ast}}} \right)^2+\left(\frac{cos\theta_{bW} - 0.23}{\sigma_{cos\theta_{bW}}} \right)^2
\label{eq:qual}
\end{equation}
In this equation $m_{cand.}$ and  $E_{cand.}$ are invariant mass and energy of the $t$ quark candidate decaying hadronically, respectively, and $m_{\tpq}$ and $E_{beam}$ are input $t$ mass and the beam energy of 250\,GeV.  Beyond that it introduces the momentum of the $b$~quark jet in the centre-of-mass frame of the $t$ quark, $p_{b}^{\ast}$, which has a defined value of $68\,\GeV$, and the angle between the $b$~quark and the $W$~boson. The measured values are compared with the expected ones and the denominator is the width of the measured distributions. Distribution of latter two observables are shown in Fig.~\ref{fig:pstarwb}. Note, that the figure shows separately good and badly reconstructed events. This is explained in Sec.~\ref{sec:afb}.
Further cuts on jet thrust $T<0.9$ and on the hadronic mass of the final state $180 < m_{had.} < 420\,\GeV$ are applied. In addition the mass windows for the reconstructed $\Wboson$-boson and $\tpq$-quark are chosen to $50<m_{\Wboson}<250$\,GeV and $120 < m_{\tpq}<270$\,GeV.

The entire selection retains 54.4\% signal events for the configuration  $\pem, \pep =-0.8,+0.3$ and 55.9\% for the configuration $\pem, \pep =+0.8,-0.3$. The fraction of (distinguishable) background events in the final sample is between 7.2\% for $\pem, \pep =+0.8,-0.3$ and 9\% for $\pem, \pep =-0.8,+0.3$. This background is predominantly composed by events from non-semileptonic decays of the $\ttbar$ pair.

Taking into account the determined efficiencies and using the born level cross given section in Table~\ref{tab:procs} together with Eq.~\ref{eq:tot-cross} a statistical uncertainty of 0.47\% in case of an initial left handed electron beam and 0.63\% in case of a right handed electron beam can be estimated. In reality these number may be somewhat worse given the fact that in particular for eL the generated sample is contaminated by single top events.
\section{Measurement of the forward backward asymmetry}\label{sec:afb}

\begin{figure}[!h]
\begin{minipage}[l]{0.45\columnwidth}
\begin{center}

\begin{fmffile}{fmfkintr}
\begin{fmfgraph*}(36,27)
\fmfleft{bl}
\fmfv{label=$\bottom_{\mathrm{lep.}}$,l.d=-0.2w,l.a=0}{bl}
\fmfright{q1,q3,q2}\fmflabel{$q$}{q1} \fmflabel{$q'$}{q2} \fmflabel{$\bottom_{\mathrm{had.}}$}{q3}
\fmf{fermion}{v1,q1}
\fmf{fermion}{v1,q2}
\fmf{fermion}{v1,bl}
\fmf{fermion}{v1,q3}
\fmffreeze
\fmfshift{-0.4w,0.0w}{q3}
\fmfshift{0.1w,0.0w}{bl}
\fmfshift{-0.1w,-0.1w}{q2}
\fmfshift{-0.1w,0.1w}{q1}
\fmfshift{-0.4w,0.0w}{v1}
\end{fmfgraph*}
\end{fmffile}

\end{center}
\end{minipage}
\hfill
\begin{minipage}[l]{0.45\columnwidth}
\begin{center}
\begin{fmffile}{fmfkintl}
\begin{fmfgraph*}(36,27)
\fmfleft{bl}
\fmflabel{$\bottom_{\mathrm{lep.}}$}{bl}
\fmfright{q1,q3,q2}\fmflabel{$q$}{q1} \fmflabel{$q'$}{q2} \fmflabel{$\bottom_{\mathrm{had.}}$}{q3}
\fmf{fermion}{v1,q1}
\fmf{fermion}{v1,q2}
\fmf{fermion}{v1,bl}
\fmf{fermion}{v1,q3}
\fmffreeze
\fmfshift{0.2w,0.0w}{q3}
\fmfshift{-0.27w,-0.2w}{q2}
\fmfshift{-0.27w,0.2w}{q1}
\fmfshift{-0.07w,0.0w}{v1}
\end{fmfgraph*}
\end{fmffile}
\end{center}
\end{minipage}
\caption{\sl  In case of a $\tpq_R$ decay, the jets from the $\Wboson$ dominate the reconstruction of the polar angle of the $t$~quark. In case of a $\tpq_L$ the $\Wboson$ is practically at rest and jets from the $\bottom$~quark dominate the and reconstruction of the polar angle of the $t$~quark.}
\label{fig:ambig}
\end{figure}
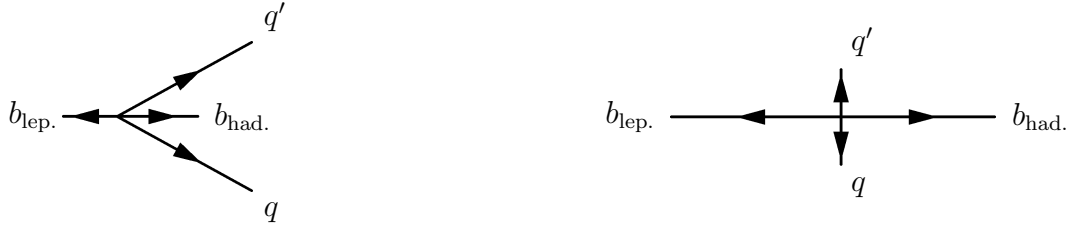

For the determination of the forward-backward asymmetry $\afbt$, the number of events in the hemispheres of the detector w.r.t. the polar angle $\theta$ of the $t$~quark is counted, i.e.
\beq
\afbt = \frac{N(\mathrm{cos}\theta>0)-N(\mathrm{cos}\theta<0)}{N(\mathrm{cos}\theta>0)+N(\mathrm{cos}\theta<0)}.
\eeqn
Here, the polar angle of the $t$ quark is calculated from the decay products in the hadronic decay branch.
The measurement of $\mathrm{cos}\theta$ depends on the correct association of the $b$~quarks to the jets of the hadronic $b$~quark decays. The analysis is carried out separately for a left-handed polarised electron beam and for a right handed polarised beam. Therefore, two different situations have to be distinguished, see also Fig.~\ref{fig:ambig}: 
\begin{itemize}
\item In case of a {\em right}-handed electron beam the sample is expected to be enriched with $\tpq$-quarks with {\em right}-handed helicity~\cite{Parke:1996pr}. Due to the $V-A$ structure of the standard model an energetic $\Wboson$~boson is emitted into the flight direction of the $\tpq$-quark. The $W$~boson decays into two energetic jets.
The $\bottom$~quark from the decay of the $\tpq$~quark are comparatively soft. Therefore, the direction of the  $\tpq$~quark is essentially reconstructed from the direction of the energetic jets from the $W$~boson decay. This scenario is thus insensitive towards a wrong association of the jet from the $b$~quark decay to the jets from the $W$~boson decay

\item In case of a {\em left}-handed electron beam the sample is enriched with $\tpq$~quarks with {\em left}-handed helicity. In this case the $\Wboson$~boson is emitted opposite to the flight-direction of the $\tpq$~quark and gains therefore only little kinetic energy. In fact for a centre-of-mass energy of 500\,GeV the $\Wboson$~boson is nearly at rest. On the other hand the $b$~quarks are very energetic and will therefore dominate the reconstruction of the polar angle of the $t$~quark.
In this case a wrong association of the jets from the $W$~boson decay with that from the $\bottom$~quark can flip the reconstructed polar angle by $\pi$ giving rise to migrations in the polar angle distribution of the $\tpq$~quark. 
\end{itemize}
The explanations above apply correspondingly to polarised positron beams and $\bar{\tpq}$-quarks.

\begin{figure}[ht]
\begin{center}
\includegraphics[width=0.7\textwidth]{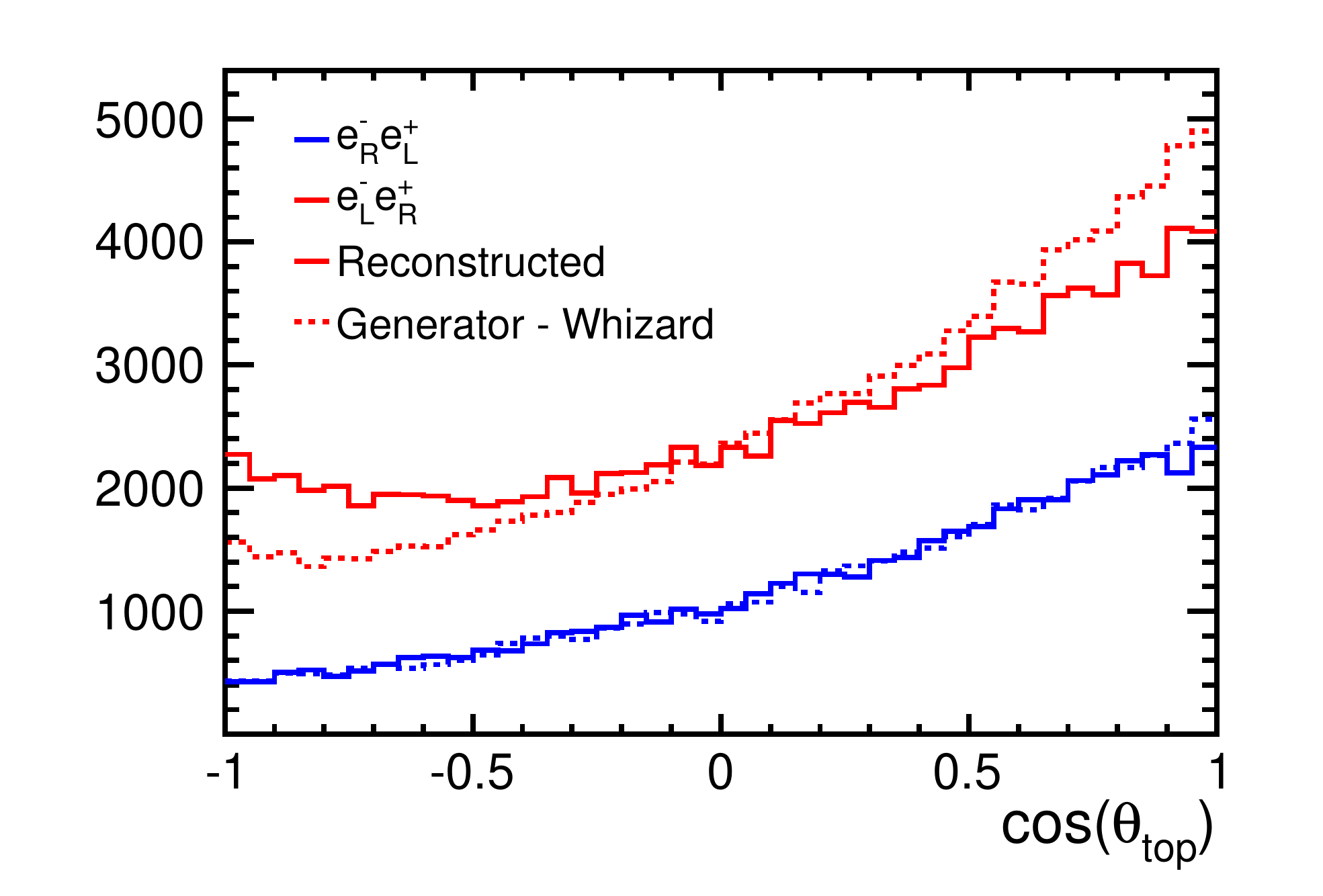}
\caption{\sl 
Reconstructed forward backward asymmetry compared with the prediction by the event generator WHIZARD for two configurations of the beam polarisations.
}
\label{fig:ambig_rec}
\end{center}
\end{figure}

The described scenarios are encountered as shown in Figure~\ref{fig:ambig_rec}. First, the reconstructed spectrum of polar angles of the $t$~quark in the case of right handed electron beams is in reasonable agreement with the generated one. On the other hand the reconstruction of $\cos{\mathrm{\theta_{\tpq}}}$ in case of left-handed $\tpq$~quarks suffers from considerable migrations. 
As discussed, the migrations are caused by a wrong association of jets stemming from $b$~quarks to jets stemming from $W$~decays. This implies that the reconstruction of observables will get deteriorated. This implication motivates to restrict the determination of $\afbt$ in case of $\pem,\pep=-1,+1$ to cleanly reconstructed events as already studied previously in~\cite{Doublet:2012wf,bib:nacho}.

The quality of the reconstructed events is estimated by the following quantity

\begin{equation}
       \chi^2=\left(\frac{\gamma_t -1.435}{\sigma_{\gamma_{t}}} \right)^2+\left(\frac{E_b^{\ast} -68}{\sigma_{E_{b}^{\ast}}} \right)^2+\left(\frac{cos\theta_{bW} - 0.26}{\sigma_{cos\theta_{bW}}} \right)^2
        \label{eq:obs}
\end{equation}

\begin{figure}[ht]
\begin{center}
\includegraphics[width=0.7\textwidth]{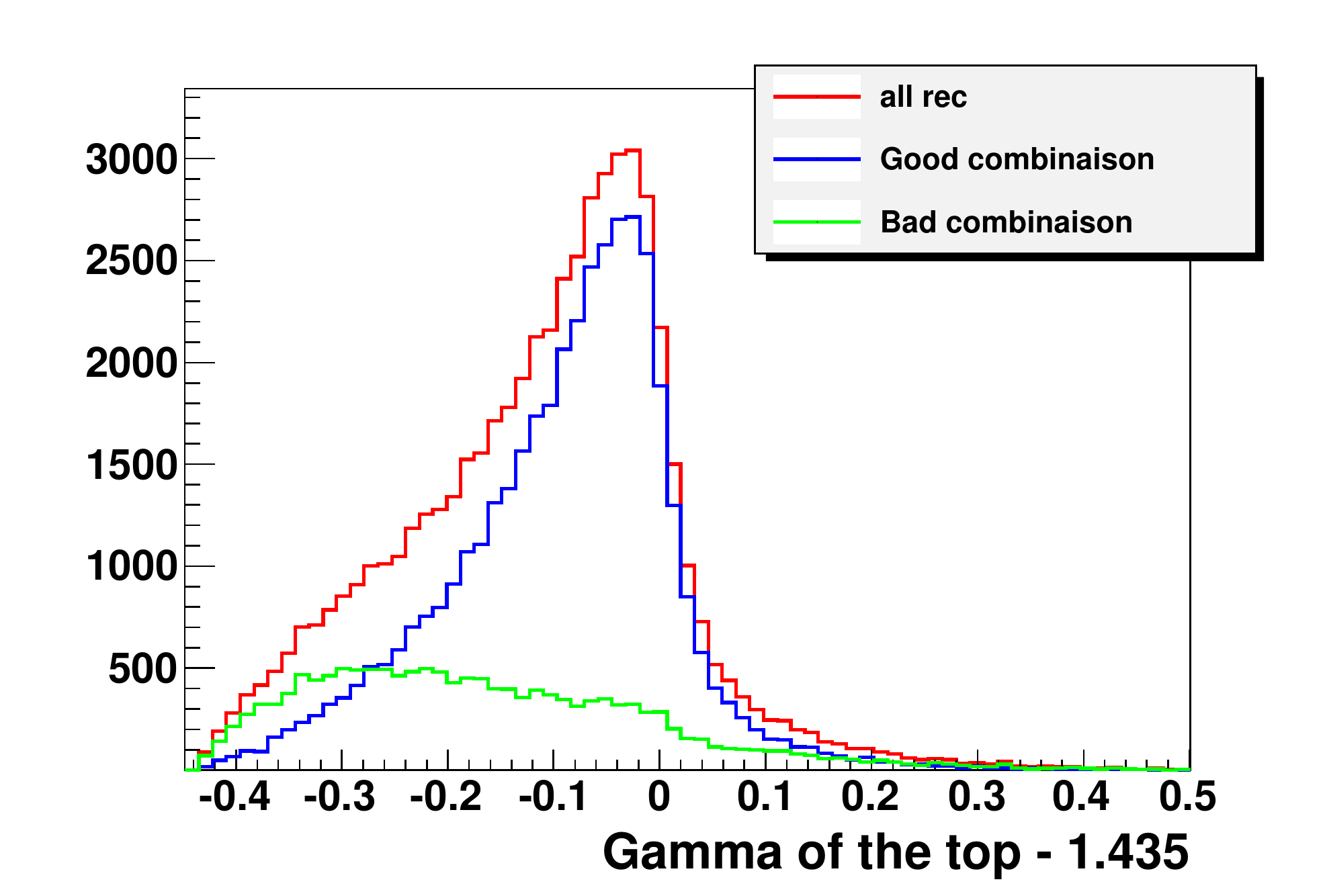}
\caption{\sl Lorentz factor of the top to define the quantity $\chi^2$, see Eq.~\ref{eq:obs}, for the selection of well reconstructed events in case of $\pem,\pep =-1,+1$ beam polarisation.
}
\label{fig:LRdistributionshad}
\end{center}
\end{figure}

The observables $p_{b}^{\ast}$ and  $\mathrm{cos}\theta_{bW}$ have already been introduced in Sec.~\ref{sec:sel}. 
The defined  $\chi^2$ comprises in addition the Lorentz factor $\gamma_t=E_t/M_t$ of the final state $t$~quark, which is shown in Figure~\ref{fig:LRdistributionshad}. 
The correct association of the of jets from $b$~quarks to that from $W$~bosons is checked with the MC truth information.
Events in which this association went wrong, labelled as {\em bad combination} in Figs.~\ref{fig:pstarwb} and~\ref{fig:LRdistributionshad}, lead to a distorted distribution in these observables.

\begin{figure}[ht]
\begin{center}
\includegraphics[width=0.7\textwidth]{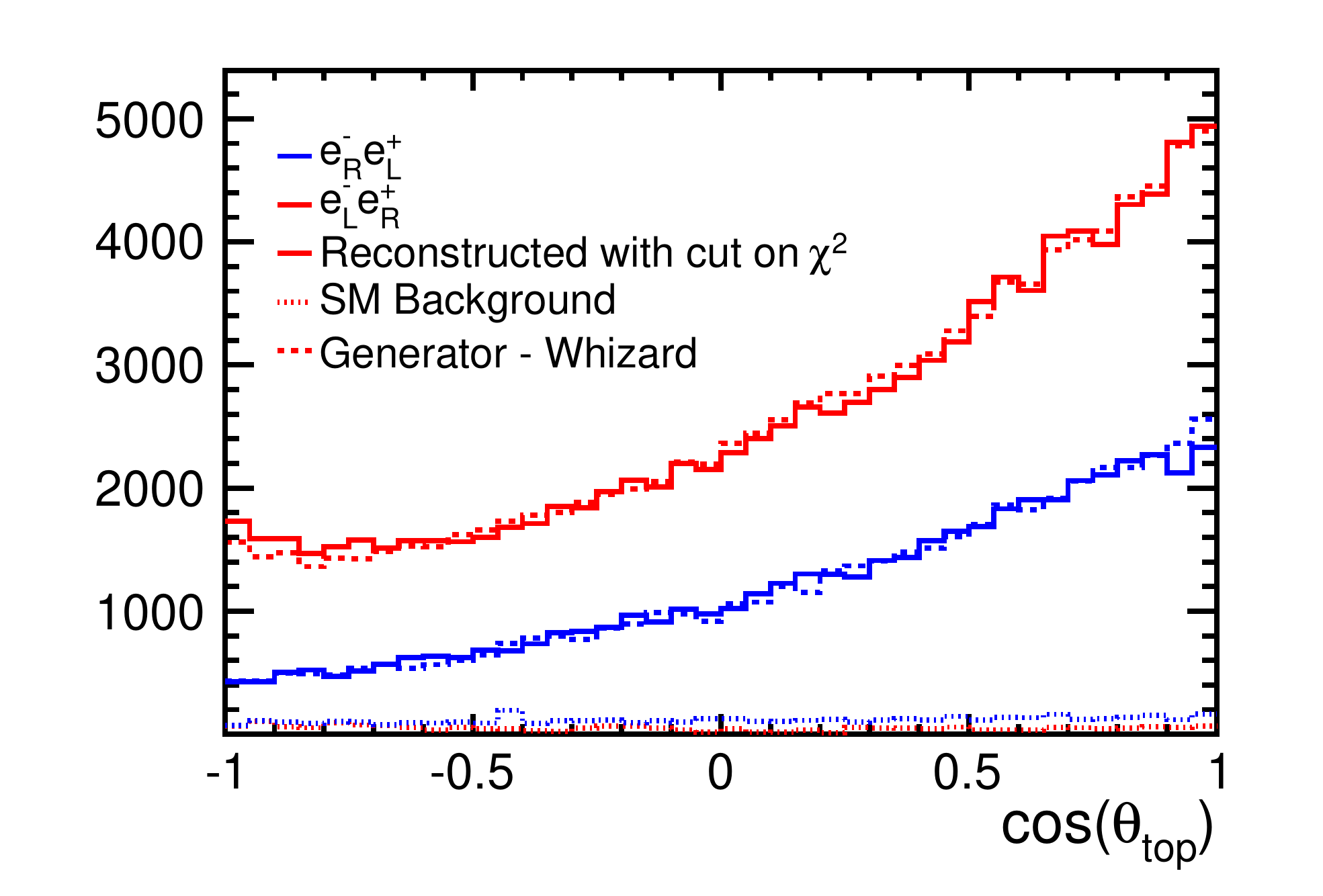}
\caption{\sl 
Reconstructed forward backward asymmetry together with residual Standard Model background compared with the prediction by the event generator WHIZARD after the application of a on $\chi^2<15$ for the beam polarisations
$P,P' =-1,+1$ as explained in the text. Note that no correction is applied for the beam polarisations $\pem,\pep =+1,-1$
}
\label{fig:ambig_reccut}
\end{center}
\end{figure}

For $\chi^2 < 15$ the reconstructed spectrum agrees very well with the generated one. 
For this cut on $\chi^2$, the reconstruction efficiency is 28.5\%. 
Fig.~\ref{fig:ambig_reccut} demonstrates the improved agreement between the reconstructed and generated direction of the $t$~quark direction in case $\pem,\pep=-1,+1$. It shows also that the residual Standard Model background is very small, e.g. less than 2\% in case of $\pep=-1,+1$ prone to be more affected by the background. The forward-backward asymmetry $\afbt$ can be derived from these angular distributions.


The numerical results are given in Tab.~\ref{tab:resafb} and compared with the generated value. The statistical error is corrected for the 
realistic beam polarisations $\pem,\pep =\pm0.8,\mp0.3$. It shows that for the standard luminosity 
statistical precisions of better than 2\% can be expected. When selecting well reconstructed events the systematic error due to the ambiguities is expected to be significantly smaller than the statistical error.


\begin{table}[h!]
  \begin{center}
    \begin{tabular}{|c|c|c|c|}
      \hline
      $\pem, \pep$ & $(\afbt)_{gen.}$& $\afbt$ & $(\delta_{\afb} / \afb)_{stat.}$ [\%]\\ \hline
      $-1,+1$& 0.339 & 0.326 &1.8 (for $\pem,\pep =-0.8,+0.3$)\\ \hline
      $+1,-1$& 0.432 & 0.420 & 1.3 (for $\pem,\pep =+0.8,-0.3$)\\ \hline
    \end{tabular}
  \end{center}
  \caption{\sl Statistical precisions expected for  $\afbt$ for different beam polarisations.}
  \label{tab:resafb}
\end{table}

\section{Determination of the slope of the helicity angle distribution}
The helicity approach has been suggested for top studies at Tevatron~\cite{Berger:2012nw}. In the rest system of the 
$t$~quark, the angle of the lepton from the $W$~boson is distributed like:

\begin{equation}\label{eq:h}
        ~\frac{1}{\Gamma}  \frac{d\Gamma}{dcos\theta_{hel}} = \frac{1+\lambda_t \mathrm{cos}\theta_{hel}}{2} = \frac{1}{2} +(2 F_R - 1)\frac{\mathrm{cos}\theta_{hel}}{2} $$\\$$ \lambda_t=1~ \mathrm{for}~ t_R~~~\lambda_t=-1~\mathrm{for} ~t_L 
\end{equation}
This angular distribution is therefore linear and very contrasted between $t_L$ and $t_R$. 
In practice there will be a mixture of $t_R$ and $t_L$ (beware that here $L$ and $R$ mean left and 
right handed helicities) and  $\lhel$ will have a value between -1 and +1 depending on the composition of 
the $t$~quark sample. 


According to~\cite{Berger:2012nw}, the angle $\thel$ is measured in the rest frame of the $t$~quark with the $z$-axis defined by the direction of motion of the $t$~quark in the laboratory. As discussed in~\cite{Parke:1996pr} this definition of  $\thel$ is not unique but some detailed investigations not reproduced in this note have shown that the choice of~\cite{Berger:2012nw} seems optimal. The observable $\cthel$ is computed from the momentum of the $t$~quark decaying semi-leptonically into a 
lepton, a $b$~quark and a neutrino. If initial state radiation effects (with the photon lost in the beam pipe) are neglected, one can simply assume energy momentum conservation. This, by means of the energy-momentum of the $t$~quark decaying hadronically, allows for deducing the energy-momentum of the $t$~quark decaying semi-leptonically. A Lorentz transformation boosts the lepton into the rest system of the $t$~quark. This should give a very precise knowledge of $\cthel$. To determine the helicity angle only the angle of the lepton needs to be known. For the leptonic decays of the $\tau$~lepton, which significantly contribute to this analysis (10-15\%), the charged lepton and the $\tau$~lepton are approximately collinear and therefore the method remains valid.






\subsection{Analysis of the helicity angle distribution}

\begin{figure}[ht]
\begin{center}
\includegraphics[width=0.7\textwidth]{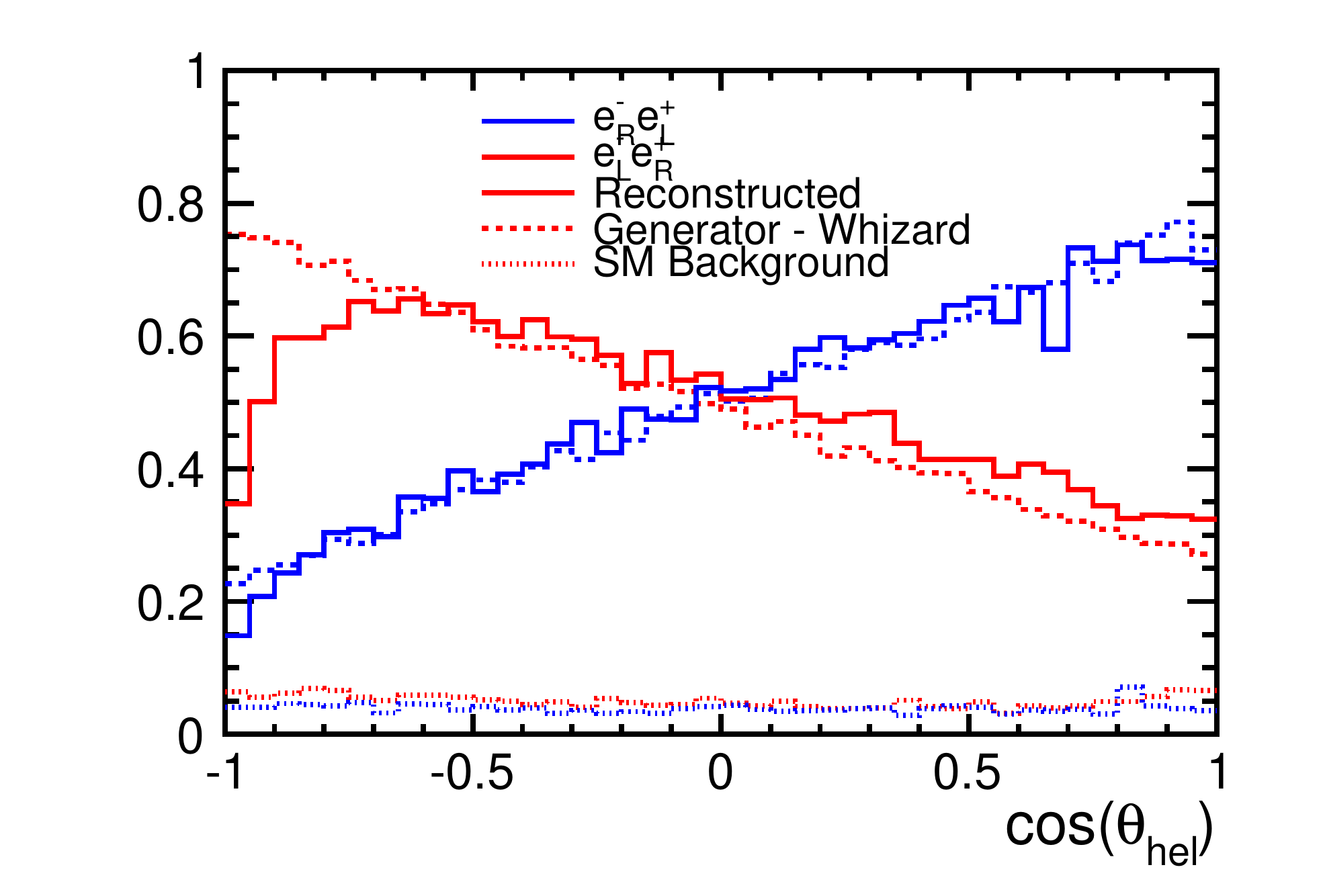}
\caption{\sl  Polar angle of the decay lepton in the rest frame of the $t$~quark.
}
\label{fig:hel_dist}
\end{center}
\end{figure}

Based on the selection introduced in Sec.~\ref{sec:sel} the angular distribution of the decay lepton in the rest frame of the $t$~quark is shown in Fig.~\ref{fig:hel_dist} together with the residual Standard Model background for fully polarised beams. The background is small relative to the signal and to a good approximation flat. It has therefore only a minor influence on the slope of the signal distribution and will be neglected in the following.

The distribution exhibits a drop in reconstructed events towards $\cthel=-1$. This drop can be explained by the
event selection which suppresses leptons with small energies. Outside this region and in contrast to e.g. the forward-backward asymmetry the reconstructed angular distribution agrees very well with the generated one. This means that this observable suffers much less from the migration effect described in Sec.~\ref{sec:afb}. It is therefore not necessary to tighten the selection in the same way as for $\afbt$. The reason for the bigger robustness of the angular distribution can be explained by kinematics.

As outlined in Sec.~\ref{sec:afb} the migrations described there are provoked mainly by longitudinally polarised, {\em soft} $W$~bosons from the decay of left handed $t$~quarks. The 
$W_L$~boson decay proportional to $\mathrm{sin}^2\theta$. Therefore any boost into the rest frame of the top leads predominantly to leptons with $\cthel<0$.  

The parameter $\lhel$ can be derived from the slope of the helicity angle distribution that is obtained by a fit to the linear part of the angular distribution in the range $\cthel=[-0.6,0.9]$ for $\pem,\pep=-1,+1$ and $\cthel=[-0.9,0.9]$ for $\pem,\pep=+1,-1$  The results are summarised in Table~\ref{tab:lhel-res} for the two initial beam polarisations $\pem=\pm 1$ and $\pep=\mp1$ and the statistical error is given for $\pem,\pep=\mp0.8,\pm0.3$. The results are compared with the values of $\lhel$ as obtained for the generated sample. A quarter of the shift between the generated and the reconstructed value is taken into account for the systematic error of the measurement.  The result changes by about 1\% when 
changing the fit range to $\cthel=[(-0.4,0.5),0.9]$ for $\pem,\pep=-1,+1$. The errors on the slope from the variation of the fit range and that from the difference between generated and reconstructed slope are added in quadrature.

\begin{table}[h!]
  \begin{center}
    \begin{tabular}{|c|c|c|c|c|}
      \hline
      $\pem, \pep$ & $(\lhel)_{gen.}$ & $(\lhel)_{rec.}$ & $(\delta \lhel)_{stat.}$  & $(\delta \lhel)_{syst.}$ \\ 
                               &                              &                             & for  $\pem,\pep=\mp0.8,\pm0.3$  & \\ \hline
      $-1,+1$  &-0.484 & -0.437 &0.011 & 0.013 \\ \hline
      $+1,-1$& 0.547 & 0.534 & 0.013 & 0.006 \\ \hline
    \end{tabular}
  \end{center}
  \caption{\sl Results on $\lhel$ derived from the slope of the helicity angle distribution with errors for different beam polarisations at the ILC.}
  \label{tab:lhel-res}
\end{table}

\section{Discussion of systematic uncertainties}

In the previous sections measurements of either cross sections or asymmetries have been presented. This section makes an attempt to identify and quantify systematic uncertainties,
which may influence the precision measurements. 

\begin{itemize}
\item \underline{Luminosity:} The luminosity is a critical parameter for cross section measurements only. The luminosity can be controlled to 0.1\%~\cite{Rimbault:2007zz}. 
\item \underline{Polarisation:} The polarisation is a critical parameter for all analyses. It enters directly the cross section measurements. The studies
for the DBD using $W$ pair production~\cite{bib:dbd:rosca} lead to an uncertainty of 0.1\% for the polarisation of the electron beam and to an uncertainty of 0.35\% for the polarisation of the positron beam. This translates into an uncertainty of 0.25\% on the cross section for $\pem,\pep=-0.8,+0.3$ and 0.18\% on the cross section for $\pem,\pep=+0.8,-0.3$ 
The uncertainty on the polarisation can be neglected with respect to the statistical uncertainty for $\afbt$ and $\lhel$.
\item \underline{Beamstrahlung and beam energy spread:} The mutual influence of the electromagnetic fields of the colliding bunches provokes radiation of photons known as {\em Beamstrahlung}. This Beamstrahlung modulates the luminosity spectrum, i.e. moves particles from the nominal energy to smaller energies. At the ILC for centre- about 60\% of the particles are expected to have 99\% or more of the nominal energy~\cite{bib:ilc-tdr-dbd}. The beam energy spread, i.e. the RMS of this main luminosity peak is 124\,MeV for the electron beam and 70\,MeV for the positron beam~\cite{bib:ilc-tdr-dbd}. Both effects play a role at the $\ttbar$ threshold~\cite{Seidel:2013sqa} and can be neglected at energies well above this threshold.

\item \underline{Experimental uncertainties in top quark reconstruction:} It has been shown in Sec.~\ref{sec:afb} that migrations have to be taken into account for the measurement of  $\afbt$, in particular for the
polarisations $\pem,\pep=-0.8,+0.3$. These migrations are reduced by stringent requirements on the event selection using a $\chi^2$ analysis. This in turn leads to a penalty in the efficiency. 
The success of the method depends in addition on a very sharp measurement of the variables used for the $\chi^2$ analysis. 
We expect that these ambiguities can be (partially) eliminated by an event-by-event determination of the charge of the $b$~quark from the $t$~decay.
\item \underline{Other experimental effects:} There is a number of other experimental effects such as acceptance, uncertainties of the $b$~tagging or the influence of passive detector material. The LEP experiments quote a systematic uncertainty on $R_b$ of 0.2\% a value which may serve as a guide line for values to be expected at the ILC, which on the 
other hand will benefit from far superior detector resolution and $b$~tagging capabilities. 
\item \underline{Theory:} 
The uncertainties of today's state-of-the-art calculations is discussed in
Section~\ref{sec:theory}. The uncertainties in the QCD corrections to
the total cross section and $A_{FB}$ are of the same order as the experimental 
uncertainties. Two-loop electro-weak corrections are required to perform
the extraction.
\item \underline{Single-top production:} Single top production at the LC in 
association with a $W$~boson and bottom quark (through $WW^{\ast}$ production) 
leads to the same final state as $t$~quark pair production. It forms a 
sizable contribution to the six-fermion final state and must be taken 
into account in a realistic experimental strategy. 
This is left for a future study. 
\item \underline{Beyond Standard Model Physics:} Possible BSM effects may affect the various components of the background, in particular the $\ttbar$ induced background. This will therefore require a careful iterative procedure with tuning of our generators. This procedure seems feasible without a significant loss of accuracy.
\end{itemize}

As a summary it can be concluded that the total systematic uncertainties will 
not exceed the statistical uncertainties. This, however, requires an excellent 
control of a number of experimental quantities on which the results depend. 



\section{Precision of Form Factors}

\begin{figure}
\centering
\includegraphics[width=\columnwidth]{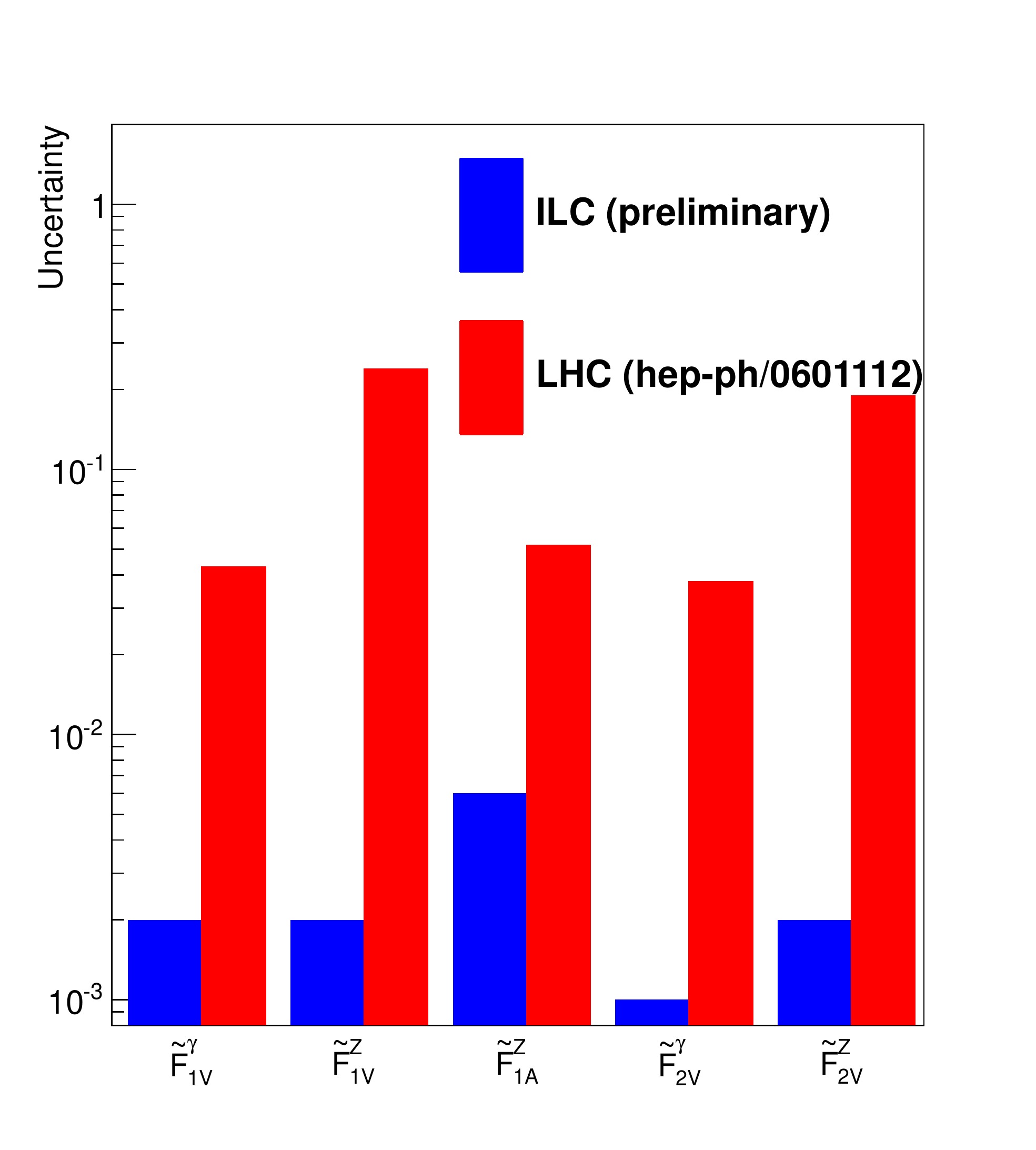}
\caption{\sl
 Comparison of statistical precisions on $CP$ conserving form factors
 expected at the LHC, taken from~\cite{Juste:2006sv} and at the ILC. 
The LHC results assume an integrated luminosity of $\mathcal{L}=300~\invfb$.
The results for ILC assume an integrated luminosity of $\mathcal{L}=500~\invfb$ at 
$\roots=500\,\GeV$ and a beam polarisation  $\pem=\pm0.8,\pep = \mp0.3$.}
\label{fig:hel-coupl}
\end{figure}

The results on the reconstruction efficiency, $\afbt$ and  $\lhel$ presented in the previous sections are transformed into precisions on the 
form factors $\widetilde F_i$. The results are summarised in Table~\ref{tab:tab1} and Figure~\ref{fig:hel-coupl} and are compared with results of earlier studies for a linear $\epem$ collider as published in the TESLA TDR~\cite{AguilarSaavedra:2001rg} as well as with precisions obtained in a simulation study for the LHC.  Note, that in the LHC and TESLA studies only one form factor was varied at a time while in the present study two or four form factors are varied simultaneously, see Sec.~\ref{sec:intro}. From the comparison of the numbers it is justified to assume that the measurements at an electron positron collider lead to a spectacular improvement and thanks to the $\gamma/Z^0$ interference a $\epem$ collider can fix the sign the form factors. At the LHC the $t$~quark couples either to the photon or to the $Z^0$. In that case the cross section is proportional to e.g. $ (F^Z_{1V})^2 +  (F^Z_{1A})^2$. The precision expected at the LHC cannot exclude a sign flip of neither  $F^Z_{1V}$ nor of $F^Z_{1A}$. On the hand the LEP bounds can exclude a sign flip for $F^Z_{1A}$ which renders a much better precision for  $\widetilde F^Z_{1A}$ compared with $\widetilde F^Z_{1V}$. Clearly, the precisions which can be obtained at the LHC are to be revisited in the light of the real LHC data. A first result on associated production of vector boson and $\ttbar$ pairs is published in~\cite{Chatrchyan:2013qca}.

For completeness, Tab.~\ref{tab:tab2} compares sensitivities obtained in the mentioned simulation study for the LHC with the results from the TESLA TDR~\cite{AguilarSaavedra:2001rg} for $CP$ violating form factors not calculated in the present study.  

The expected high precision at a linear $\epem$ collider allow for a profound discussion of effects of new physics. The findings can be confronted with predictions in the framework of Randall-Sundrum models and/or compositeness models such as~\cite{Pomarol:2008bh,Djouadi:2006rk,Hosotani:2005nz,Cui:2010ds,Carena:2006bn,Grojean:2013qca} or Little Higgs models as e.g.~\cite{Berger:2005ht}. All these models entail deviations from the Standard Model values of the $t$~quark couplings to the $Z^0$ boson that will be measurable at the ILC. The interpretation of the results presented in this article in terms of the cited and maybe other models is in preparation and left for a future publication. {\em Comments and contributions from theory groups are highly welcome.}



\begin{table}[t]
\begin{center}
\begin{footnotesize}
\begin{tabular}{|ccccc|}
\hline 
 Coupling & SM value & LHC~\protect\cite{Juste:2006sv}  & $e^+e^-$~\protect\cite{AguilarSaavedra:2001rg}& 
           $e^+e^-$[{\em ILC DBD}]\\
& & $\mathcal{L}=300~\invfb$ &  $\mathcal{L}=300~\invfb$  & $\mathcal{L}=500~\invfb$\\
& &                                                &     $\pem,\pep=-0.8,0$                                        & $\pem,\pep =\pm0.8,\mp0.3$\\
\hline
$\Delta\widetilde F^\gamma_{1V}$ & 
0.66 &
$\begin{matrix} +0.043 \\[-4pt] -0.041\end{matrix}$ & 
$\begin{matrix} - \\[-4pt] - \end{matrix}$&   
$\begin{matrix} +0.002 \\[-4pt] -0.002 \end{matrix}$
\\
$\Delta\widetilde F^Z_{1V}$ &
0.23 &
$\begin{matrix} +0.240 \\[-4pt]  -0.620\end{matrix}$ & 
$\begin{matrix} +0.004 \\[-4pt] -0.004\end{matrix}$ &
$\begin{matrix} +0.002 \\[-4pt] -0.002\end{matrix}$  
\\
$\Delta\widetilde F^Z_{1A}$ &
-0.59 &
$\begin{matrix} +0.052 \\[-4pt]  -0.060\end{matrix}$ & 
$\begin{matrix} +0.009 \\[-4pt] -0.013\end{matrix}$ &
$\begin{matrix} +0.006 \\[-4pt] -0.006\end{matrix}$ 
\\
$\Delta\widetilde F^\gamma_{2V}$ & 
0.015 &
$\begin{matrix} +0.038 \\[-4pt] -0.035\end{matrix}$ & 
$\begin{matrix} +0.004 \\[-4pt] -0.004\end{matrix}$ &   
$\begin{matrix} +0.001 \\[-4pt] -0.001\end{matrix}$
\\
$\Delta\widetilde F^Z_{2V}$ & 
0.018 &
$\begin{matrix} +0.270 \\[-4pt] -0.190\end{matrix}$ & 
$\begin{matrix} +0.004 \\[-4pt] -0.004\end{matrix}$ & 
 $\begin{matrix} +0.002 \\[-4pt] -0.002\end{matrix}$  
\\
\hline

\end{tabular}
\end{footnotesize}
\caption{\sl Sensitivities achievable at $68.3\%$ CL for $CP$ conserving form factors $\widetilde F^X_{1V,A}$ and $\widetilde F^X_{2V}$ defined in Eq.~\ref{eq:snow} at the LHC and at linear $\epem$ colliders.  The assumed luminosity samples and, for $\epem$ colliders, the beam polarisation, are indicated.  In the LHC studies and in earlier studies for a linear $\epem$ collider as published in the TESLA TDR~\cite{AguilarSaavedra:2001rg} study, only one coupling at a time is allowed to deviate from its Standard Model value.  In the present study, denoted as {\em ILC DBD}, either the four form factors $\widetilde F_1$ or the two form factors $\widetilde F_2$ are allowed to vary independently. The sensitivities are based on statistical errors only.}
\label{tab:tab1}
\end{center}
\end{table}

\begin{table}[ht]
\begin{center}
\begin{footnotesize}
\begin{tabular}{|ccc|}
\hline 
 Coupling & LHC~\protect\cite{Juste:2006sv}  &
           $e^+e^-$~\protect\cite{AguilarSaavedra:2001rg}\\
 & $\mathcal{L}=300~\invfb$ & 
$\mathcal{L}=300~\invfb$\\
& & $\pem,\pep =-0.8,0$\\
\hline
$\Delta {\mbox Re}\, \widetilde F_{2A}^{\gamma}$& $\begin{matrix} +0.17 \\[-4pt] -0.17\end{matrix}$ &$\begin{matrix} +0.007 \\[-4pt] -0.007\end{matrix}$   \\
$\Delta {\mbox Re}\, \widetilde F_{2A}^{Z}$& $\begin{matrix} +0.35 \\[-4pt] -0.35\end{matrix}$ &$\begin{matrix} +0.008 \\[-4pt] -0.008\end{matrix}$  \\
$\Delta {\mbox Im}\, \widetilde F_{2A}^{\gamma}$& $\begin{matrix} +0.17 \\[-4pt] -0.17\end{matrix}$  &$\begin{matrix} +0.008 \\[-4pt] -0.008\end{matrix}$  \\
$\Delta {\mbox Im}\, \widetilde F_{2A}^{Z}$& $\begin{matrix} +0.035 \\[-4pt] -0.035\end{matrix}$ & $\begin{matrix} +0.015 \\[-4pt] -0.015\end{matrix}$ \\
\hline
\end{tabular}
\end{footnotesize}
\caption{\sl Sensitivities achievable at $68.3\%$ CL for the top quark
magnetic and electric dipole form factors  $\widetilde F^V_{2A}$ defined in Eq.~\ref{eq:snow}, at the LHC and at 
for a linear $\epem$ collider as published in the TESLA TDR~\cite{AguilarSaavedra:2001rg}.  The assumed luminosity samples and, for TESLA, beam polarisation, are indicated. In the LHC study and in the TESLA study only one coupling at a time is 
allowed to deviate from its Standard Model value. The sensitivities are based on statistical errors only} 
\label{tab:tab2}
\end{center}
\end{table}

\section{Summary and outlook}

This article presents a comprehensive analysis of $\ttbar$~quark production using the semi-leptonic decay channel. Results are given for a centre-of-mass energy of $\roots = 500\,\GeV$ and an integrated luminosity of $\mathcal{L}=500\,\invfb$ shared equally between the beam polarisations. $P=\pm 0.8$ and $P'=\mp 0.3$.  

Semi-leptonic events, including those with $\tau$ leptons in the final state can be selected with an efficiency of about 55\%.
The cross section of the semi-leptonic channel of $\ttbar$~quark production can therefore be measured to a statistical precision of about 0.5\%. The second observable is the forward-backward asymmetry $\afbt$. It was shown that in particular for predominantly left handed polarisation of the initial electron beam the $V-A$ structure leads to migrations, which distort the theoretical expected $\afbt$. These migrations can be remedied by tightening the selection criteria of the events. Taking into account this correction the forward-backward asymmetry can be determined to a precision of better than 2\% for both beam polarisations. Finally, the study introduced the slope of the helicity angle distribution, which is a new observable for ILC studies. This observable measures the fraction of $t$~quarks of a given helicity in the event sample. This variable is very robust against e.g. the migration effects and can be measured to a precision of about 4\%. 

The observables together with the unique feature of the ILC to provide polarised beams allow for a largely unbiased disentangling of the individual couplings of the $t$~quark to the $Z^0$~boson and the photon. These couplings can be measured with high precision at the ILC and, when referring to the results in~\cite{Juste:2006sv}, always more than one order of magnitude better than it will be possible at the LHC with $\mathcal{L}=300\,\invfb$. A study of form factors using real LHC data is however eagerly awaited.

The precision as obtained in the present study for the ILC would allow for the verification of a great number of models for physics beyond the Standard Model. Examples for these models are extra dimensions and compositeness. The results obtained here constitute therefore a perfect basis for discussions with theoretical groups. 
In extension of~\cite{AguilarSaavedra:2001rg} a future analysis will disentangle the $CP$ violating form factors. 

It has to be noted that the results contain only partially experimental systematical errors. These will have to be estimated in future studies. From the achieved precision it is mandatory that systematics are controlled to the 1\% level or better in particular for the measurement of the cross section.

\bibliographystyle{utphys_mod}
\begin{footnotesize}
\bibliography{tt-note}
\end{footnotesize}

\end{document}